# A *general moment* NRIXS approach to the determination of equilibrium Fe isotopic fractionation factors: application to goethite and jarosite

N. Dauphas[a,*], M. Roskosz[b], E.E. Alp[c], D.C. Golden[d], C.K. Sio[a],
F.L.H. Tissot[a], M. Hu[c], J. Zhao[c], L. Gao[c], R.V. Morris[e]

[a]Origins Laboratory, Department of the Geophysical Sciences and Enrico Fermi Institute, The University of Chicago, 5734 South Ellis Avenue, Chicago IL 60637, USA

[b]Unité Matériaux et Transormations, Université de Lille 1, CNRS UMR 8207, 69655 Villeneuve d'Ascq, France

[c]Advanced Photon Source, Argonne National Laboratory, 9700 South Cass Avenue, Argonne, IL 60439, USA

[d]Engineering and Science Contract Group-Hamilton Sundstrand, Mail Code JE23, Houston, TX 77058, USA

[e]NASA Johnson Space Center, Houston, TX, USA

*Corresponding author
*Email address*: dauphas@uchicago.edu

**Abstract**

The equilibrium Fe isotopic fractionation factors (1,000×lnβ) of goethite and jarosite have considerable importance for interpreting Fe isotope variations in low-temperature aqueous systems on Earth and possibly Mars; in the context of future sample return missions. We measured the β-factors of goethite FeO(OH), potassium-jarosite $KFe_3(SO_4)_2(OH)_6$, and hydronium-jarosite $(H_3O)Fe_3(SO_4)_2(OH)_6$, by Nuclear Resonant Inelastic X-Ray Scattering (NRIXS, also known as Nuclear Resonance Vibrational Spectroscopy –NRVS- or Nuclear Inelastic Scattering -NIS) at the Advanced Photon Source. These measurements were made on synthetic minerals enriched in $^{57}Fe$. A new method (*i.e.*, the *general moment* approach) is presented to calculate β-factors from the moments of the NRIXS spectrum *S(E)*. The first term in the moment expansion controls iron isotopic fractionation at high temperature and corresponds to the mean force constant of the iron bonds, a quantity that is readily measured and often reported in NRIXS studies. The mean force constants of goethite, potassium-jarosite, and hydronium-jarosite are 314±14, 264±12, and 310±14 N/m, respectively (uncertainties include statistical and systematic errors). The *general moment* approach gives $^{56}Fe/^{54}Fe$ β-factors of 9.7, 8.3, and 9.5 ‰ at 22 °C for these minerals. The β-factor of goethite measured by NRIXS is larger than that estimated by combining results from laboratory exchange experiments and calculations based on electronic structure theory. Similar issues have been identified previously for other pairs of mineral-aqueous species, which could reflect inadequacies of approaches based on electronic structure theory to calculate absolute β-factors (differences in β-factors between aqueous species may be more accurate) or failure of laboratory experiments to measure mineral-fluid equilibrium isotopic fractionation at low temperature. We apply the force constant approach to published NRIXS data and report 1,000lnβ for important Fe-bearing phases of geological and biochemical relevance such as myoglobin, cytochrome f, pyroxene, metal, troilite, chalcopyrite, hematite, and magnetite.

## 1. Introduction

In near-neutral oxic conditions, oxidation of ferrous iron ($Fe^{2+}$) can take place very rapidly to produce ferric iron ($Fe^{3+}$) that rapidly precipitates as insoluble forms such as ferric oxyhydroxide, explaining the widespread occurrence of these minerals in low temperature aqueous systems (STUMM and MORGAN, 1996). Goethite and hematite are the most abundant Fe-bearing oxides present at Earth's surface, where they are ubiquitous in soils as well as lacustrine and marine sediments. At low temperature (<100 °C), goethite is thermodynamically favored over most iron oxides but other parameters such as reaction kinetics, grain size, water activity, and element substitution can influence the nature of the minerals that form (CORNELL and SCHWERTMANN, 2003). On Mars, acidic conditions resulted in precipitation of the sulfate mineral jarosite (KLINGELHÖFER et al., 2004) at Meridiani Planum, and goethite has been detected in outcrop rocks of the Columbia Hills at Gusev Crater (MORRIS et al., 2006, 2008). Goethite may also be a precursor to hematite associated with jarosite at Meridiani Planum (TOSCA et al., 2005; ZOLOTOV and SHOCK, 2005). Similar mineral assemblages to Martian outcrops have been found on Earth in acid mine drainage environments, in natural pyritic ore bodies (FERNANDEZ-REMOLAR et al., 2005; NORDSTROM DARRELL et al., 1979) and in volcanic edifices near fumaroles (MORRIS et al., 2005).

The question of the biogeochemical transformations of iron-bearing minerals and their relationships to dissolved species can be investigated by measuring iron isotope variations in rocks, minerals and fluids (DAUPHAS and ROUXEL, 2006; JOHNSON et al., 2008). For example, coupled carbon and iron isotope studies of banded iron formations have revealed the antiquity of microbial iron respiration in sediments (CRADDOCK and DAUPHAS, 2011; HEIMANN et al., 2010). However, a major difficulty for interpreting iron isotope variations is that most equilibrium fractionation factors have not been rigorously determined. Several approaches can be used to estimate equilibrium Fe isotope fractionation factors:

(1) Products of laboratory experiments can be directly measured (BEARD et al., 2010; POITRASSON et al., 2009; SAUNIER et al., 2011; SCHUESSLER et al., 2007; SHAHAR et al., 2008; SKULAN et al., 2002; WELCH et al., 2003; WU et al., 2011). The main difficulties in this approach are to assess the degree of equilibration and to monitor and correct for all kinetic isotope effects that could affect the results.

(2) Calculations based on electronic structure theory and other computational methods provide estimates of reduced partition function ratios or β-factors (ANBAR et al., 2005; BLANCHARD et al., 2009; DOMAGAL-GOLDMAN and KUBICKI, 2008; DOMAGAL-GOLDMAN et al., 2009; HILL and SCHAUBLE, 2008; HILL et al., 2009; OTTONELLO and ZUCCOLINI, 2008, 2009; RUSTAD and YIN, 2009; SCHAUBLE et al., 2001). The accuracy of these methods can be difficult to assess and the results are variable depending on the assumptions that are made and the parameters that are used.

(3) The vibrational properties of iron in solids can be investigated by using nuclear resonance inelastic X-ray scattering (NRIXS), from which β-factors can be derived (POLYAKOV, 2009; POLYAKOV et al., 2007; POLYAKOV and SOULTANOV, 2011).

This last method is relatively new in isotope geochemistry (POLYAKOV et al., 2005a,b). Initially, Polyakov and coworkers used Mössbauer spectroscopy to calculate iron β-factors and to identify the parameters that affect iron isotopic fractionation such as oxidation state (POLYAKOV, 1997; POLYAKOV and MINEEV, 2000). While this method is theoretically sound, the determination of second-order Doppler shifts is difficult using conventional Mössbauer spectroscopy measurements. The positions of the absorption lines shift as a function of temperature for a variety of reasons, which are additive. These include second order Doppler shift, which is determined by kinetic energy and (chemical) isomer shift. Additional complications arise if there is a distribution of hyperfine field parameters caused by changes of chemical or magnetic nature. Carefully crafted experiments are needed to disentangle these effects reliably and reproducibly. For example, Polyakov and Mineev (2000) predicted the equilibrium fractionation factor of magnetite, based on conventional Mössbauer data from De Grave et al. (1993), that was later proven to be inaccurate when mineral equilibration experiments were performed (SHAHAR et al., 2008) and new conventional Mössbauer and NRIXS data became available (POLYAKOV et al., 2007 and references therein). NRIXS is a far superior method than Mössbauer spectroscopy that uses a synchrotron X-ray source to probe the vibrational properties of iron atoms in the solid lattice. Polyakov and coworkers used the kinetic energy derived from the partial phonon density of state (PDOS) of the Fe sublattice to estimate β-factors. In this paper, we outline a superior approach based on the determination of the moments of the NRIXS spectrum $S(E)$. At high temperature, this approach reduces to determining the iron force constant, which is given by the third moment of the NRIXS spectrum.

NRIXS data for synthetic powders of $^{57}$Fe-enriched goethite, K-jarosite, and H$_3$O-jarosite are reported and the geochemical implications of these measurements are presented.

**2. Materials and Methods**

*Synthesis of $^{57}$Fe-enriched minerals*

Because NRIXS is only sensitive to $^{57}$Fe among all Fe isotopes (its natural abundance is 2.119 %) and synchrotron beam-time is valuable, samples enriched in $^{57}$Fe were prepared in order to minimize beam time. The starting material consists of $^{57}$Fe-rich metal (95 %) purchased from Cambridge Isotope Laboratories, Inc. Details of the mineral synthesis have been published previously (GOLDEN et al., 2008) and are briefly described below. For the synthesis of hydronium-jarosite, 100 mg of $^{57}$Fe powder was transferred into a clean 23-mL Teflon-lined Parr reaction vessel. Approximately 145 μL of 18M H$_2$SO$_4$, 40 μL of HClO$_4$ (oxidant), and 2 mL of H$_2$O were added to this vessel, which was closed airtight and heated for 17 h in an oven at 140 °C. The reaction vessel was cooled to room temperature in a freezer before opening. The pH was recorded (~0.6) and it was brought up to ~1 (actual pH = 1.08) by adding appropriate amounts of Mg-hydroxycarbonate (~ 1 mg). The vessel was closed airtight and was heated a second time at 140 °C for 17 h. The run-product was washed in distilled water by centrifugation and decantation in a high-speed centrifuge. The final product was freeze-dried and stored until further use. Potassium-jarosite was prepared in a similar fashion except that before the second heating step, 54 mg of K$_2$SO$_4$ was added and the pH was adjusted to ~2.1-2.75 (actual value = 2.72) by adding Mg-hydroxycarbonate. Goethite was precipitated by the same protocol as that used for the synthesis of hydronium-jarosite except that the pH was adjusted to 2.5 instead of 1. It is worth noting that the addition of Mg-hydroxycarbonate for pH adjustment does not affect jarosite formation because Mg only forms soluble sulfates that were efficiently removed during the washing step. The nature of all the minerals synthesized was checked by X-ray powder diffraction. The particle sizes (equivalent sphere diameters of coherent domain sizes) were determined using the peak broadening of Rietveld refined XRD (JADE software package, Materials Data Inc.). The particle sizes are 19.8 nm for goethite and 223.0 nm for K-Jarosite.

*Principles of NRIXS spectroscopy and experimental setup*

Nuclear Resonant Inelastic X-Ray Scattering (NRIXS) is a recently established spectroscopic technique that allows one to probe the vibrational properties of certain elements in solids (CHUMAKOV and STURHAHN, 1999; STURHAHN et al., 1995; Seto et al. 1995). Its use in Earth sciences has focused on high-pressure applications to determine seismic velocities and phonon density of states of minerals relevant to Earth's deep mantle and core (STURHAHN and JACKSON, 2007). More recently, Polyakov and coworkers have shown that NRIXS could be used in stable isotope geochemistry to determine equilibrium fractionation between minerals (POLYAKOV, 2009; POLYAKOV et al., 2007; POLYAKOV et al., 2005a,b; POLYAKOV and SOULTANOV, 2011). The NRIXS method, like conventional Mössbauer spectroscopy, relies on the fact that $^{57}$Fe possesses a low-lying nuclear excited state at 14.4125 keV that can be populated by X-ray photons of the appropriate energy. Conventional Mössbauer Spectrometers are not sufficiently tunable to cover the phonon spectral range, which is typically within ± 80 meV. Typical Mossbauer drives operate at ± 100 mm/sec, corresponding to ± 500 neV (for $^{57}$Fe transition). Sophisticated apparatus have extended this range to 2.6 meV, but for very limited applications (Röhlsberger et al. 1997). In NRIXS, the photon source is a bright X-ray beam produced by a synchrotron. The primary beam is pulsed and has a broad energy spectrum, from 1 to 1,000 keV. It passes through two monochomators that reduce this energy spread to 1 meV (Fig. 1). These high-resolution monochromators use several crystals in Bragg diffraction positions (TOELLNER, 2000). This monochromatic beam is then focused using X-ray mirrors to ~10 μm. The signal resulting from interaction of the incident beam with $^{57}$Fe in the sample is measured at an angle on the same side as the incident beam. The detector is an Avalanche Photo-Diode (APD). Extensive details on NRIXS and its applications can be found in several contributions (ALP et al., 2002; CHUMAKOV and STURHAHN, 1999; STURHAHN and JACKSON, 2007; STURHAHN et al., 1995). Interactions of the incident beam with the sample are described in more detail below.

Most of the incident X-rays are scattered by electronic shells without interacting with the Fe nuclei. The signal from these photons is almost instantaneous (<$10^{-12}$ s). A small fraction of the incident X-rays can excite the $^{57}$Fe nucleus. The lifetime of this excited nucleus is 141 ns, meaning that the signal from nuclear scattering will be received with some delay. Because the primary beam is pulsed (153 ns between pulse, each of 70 ps duration), it is possible to apply a time discriminator on the incoming signal, remove the electron contribution and retain only the nuclear scattering signal. The monochromator is tunable by physically rotating the crystals to change the Bragg angle. A typical energy scan would be from -80 to +80 meV around the elastic peak at 14.4125 keV.

Recoilless elastic scattering, as in Mössbauer spectroscopy, produces part of the signal. The rest of the nuclear scattering signal is influenced by the vibrational properties of the iron lattice. Atoms in solids are in oscillating movements. These collective movements can be decomposed into normal modes of vibration that have particle-like properties known as phonons. When a photon of higher energy than the resonance energy impacts a $^{57}$Fe nucleus, part of that energy can go into exciting the nucleus while the rest can go into exciting certain modes of vibration of the solid lattice. This process is known as phonon creation. Conversely, the incident photon can have lower energy than what is required to excite $^{57}$Fe but part of the vibrational energy in the crystal lattice can fill this energy gap and allow for the transition to happen. This is known as phonon annihilation. Thus, the nuclear resonant transitions that happen off the nominal resonance energy are influenced by the phonon energy spectrum. After data processing and assuming that the interatomic potential is quadratic in the atomic displacement (harmonic approximation), it is possible to retrieve the partial phonon density of state (PDOS) (CHUMAKOV and STURHAHN, 1999; STURHAHN et al., 1995). The term "partial" refers to the fact that the phonon density of state is only relevant to the iron sub-lattice. When working on a large monocrystal, the PDOS is also projected in the sense that it only probes the solid in the direction of the incident beam. When working with fine powder or glass, this is not an issue as the material is isotropic at the scale of the incident beam.

The powdered samples of jarosite and goethite were mounted as compressed pellets into aluminum holders. The NRIXS measurements were done at beamline 3-ID-B of the Advanced Photon Source (APS) at Argonne National Laboratory. The storage ring was operated in top-up mode with 24 bunches separated by 153 ns. The sample tilt relative to the incident beam was 5° and an APD detector was positioned a few millimeters away from the sample. A second APD was positioned downstream to measure the average energy resolution (full width at half maximum) of 1.33 meV. The monochromator was tuned from -120 to +130 meV with a step size of 0.25 meV and a collection time of 5 s per step. The minerals were rich in Fe, so 1 or 2 scans were sufficient to yield high-quality data. All data were acquired at room pressure and room temperature (299±1 K). Data reduction was done with the PHOENIX software (STURHAHN, 2000) and a Mathematica script written by the authors (Appendix A).

**3. Application of NRIXS Spectroscopy to Isotope Geochemistry**

Iron possesses an isotope with a low-lying nuclear excited state, making it possible to directly probe its vibrational properties using synchrotron radiation and to calculate equilibrium isotopic fractionation factors (POLYAKOV, 2009; POLYAKOV et al., 2007; POLYAKOV and SOULTANOV, 2011). To be more precise, NRIXS allows one to derive reduced partition function ratios or β-factors, from which equilibrium isotopic fractionation between two phases A and B can be calculated by using, $\delta_A - \delta_B \simeq 1{,}000 \times (\ln\beta_A - \ln\beta_B)$. Below, we present a new approach to calculate β-factors from the moments of the NRIXS spectrum $S(E)$. The first term in the moment expansion controls iron isotopic fractionation at high temperature and corresponds to the mean force constant of the iron bonds, a quantity that is readily measured and often reported in NRIXS studies.

3.1. *General moment* approach

Iron *β*-factors can be calculated based on series expansions in moments of the PDOS of the iron kinetic energy (Polyakov et al. 2005b, 2007) or reduced partition function ratio (Appendix B). The expansion is well approximated by the three lowest order terms,

$$1{,}000 \ln \beta_{I/I^*} \simeq 1{,}000 \left(\frac{M}{M^*} - 1\right)\left(\frac{m_2^g}{8k^2T^2} - \frac{m_4^g}{480k^4T^4} + \frac{m_6^g}{20{,}160k^6T^6}\right), \quad (1)$$

where $I$ and $I^*$ are two isotopes of iron of masses $M$ and $M^*$, and the $j^{\text{th}}$ moment of the PDOS $g(E)$ is given by,

$$m_j^g = \int_0^{+\infty} E^j g(E) dE. \quad (2)$$

Polyakov et al. (2005a, b, 2007), Polyakov (2009) and Polyakov and Soultanov (2011) used this equation to calculate *β*-factors in minerals from NRIXS data. In Appendix C, we show for the first time that the even moments of $g$ can be obtained from the moments of $S$ and Eq. 1 can be rewritten as,

$$1{,}000\ln\beta_{I/I^*} = 1{,}000\left(\frac{M}{M^*} - 1\right)\frac{1}{E_r}\left[\frac{R_3^S}{8k^2T^2} - \frac{R_5^S - 10R_2^S R_3^S}{480k^4T^4} + \frac{R_7^S + 210(R_2^S)^2 R_3^S - 35R_3^S R_4^S - 21R_2^S R_5^S}{20{,}160k^6T^6}\right],$$

(3)

where the $j^{th}$ moment of $S$ centered on $E_R$ is,

$$R_j^S = \int_{-\infty}^{+\infty} S(E)(E - E_R)^j dE, \quad (4)$$

with $E_R = E_0^2/2Mc^2$ the free recoil energy (1.956 meV for the $E_0$=14.4125 keV nuclear transition of $^{57}$Fe). When raw NRIXS data are available, all the moments of $S(E)$ can be calculated and $1{,}000 \ln \beta$ can be written as $A_1/T^2 + A_2/T^4 + A_3/T^6$. It is recommended that Eq. 3 be used over Eq. 1 in future applications to Fe isotope geochemistry because the force constant estimated from $S(E)$ is less sensitive to the assumed background than that derived from $g(E)$ (provided that it is constant and the energy range is symmetric), $S(E)$ can be extrapolated beyond the energy acquisition range to account for missing $n$-phonon contributions (but not missing normal mode frequencies), and coefficients obtained using Eq. (3) have smaller statistical uncertainties than those obtained using Eq. (1) (Fig. 2).

At high temperature, the first terms in Eq.1 and 3 provide a sufficient approximation (Fig. 3 shows under which conditions this formula is valid),

$$1{,}000 \ln \beta_{I/I^*} \simeq 1{,}000\left(\frac{M}{M^*} - 1\right)\frac{m_2^g}{8k^2T^2} = 1{,}000\left(\frac{M}{M^*} - 1\right)\frac{1}{E_r}\frac{R_3^S}{8k^2T^2}. \quad (5)$$

We recognize in these expressions a term that corresponds to the average restoring force constant of the harmonic oscillators holding the element in position (LIPKIN, 1995; LIPKIN, 1999, KOHN and CHUMAKOV, 2000),

$$\langle F \rangle = \frac{M}{\hbar^2}\int_0^{+\infty} E^2 g(E)dE = \frac{M}{E_R \hbar^2}\int_{-\infty}^{+\infty}(E - E_R)^3 S(E)dE. \quad (6)$$

Therefore, Eq. 5 can be rewritten as,

$$1{,}000 \ln \beta_{I/I^*} = 1{,}000\left(\frac{1}{M^*} - \frac{1}{M}\right)\frac{\hbar^2}{8k^2T^2}\langle F \rangle, \quad (7)$$

which is a familiar formula in isotope geochemistry (Herzfeld and Teller 1938, Bigeleisen and Goeppert-Mayer 1947). The force constant in Eq. 6 corresponds to the second-order derivative of the interaction potential, which should be constant for a harmonic oscillator (LIPKIN, 1995; LIPKIN, 1999). Measuring the force constant at different temperatures offers a means of testing the possible anharmonicity of lattice vibrations (STURHAHN, 2004). The force constant calculated using the excitation function *S(E)* is less sensitive to background subtraction than that derived from *g(E)* because positive and negative terms annihilate in Eq. 6 for symmetric energy scans; $\int_{-x}^{+x} bE^3 dE = 0$ with *b* constant. Note that in reality, *b* may change with time and the scan may not be perfectly centered and symmetric, which can affect the force constant calculated from *S(E)*. Most papers reporting NRIXS data give the average force constant computed in this manner. The force constant can also be calculated from the PDOS *g(E)* but this method is less reliable as the background can affect it and the statistical uncertainties are larger (Fig. 2). Nonetheless, in properly acquired measurements, the force constant calculated using either method should agree. The energy cube or square (Eq. 6) are factors in the integrands that give the force constant. This means that even small bumps in the high-energy tails of the excitation function *S(E)* or in the PDOS g(*E*) can have sufficient weight to affect the force constant and hence the equilibrium isotopic fractionation factors. Because these tails are characterized by low counting statistics, determining mean force constants is particularly challenging, as it requires broad energy scans and long acquisition times. A potential difficulty arises when the energy scan has insufficient width and some NRIXS signal from multiple phonons at high energy is truncated. The PHOENIX software allows one to assess and correct for this possible shortcoming by calculating the multiple phonon-contribution after single phonon decomposition and using this to extrapolate *S(E)* in a physically sound manner ("force constant after refinement" in Table 2). However, this extrapolation method cannot account for normal vibration modes that would be present outside of the scanned energy range and those buried under the background.

3.2. Application to published force constant data

Below, we provide an alternative approximate approach that is particularly useful for calculating equilibrium isotope fractionation from published force constants derived from NRIXS data. The 4$^{th}$ moment of the PDOS (second term in Eq. 1) can be estimated from the 2$^{nd}$ moment if we approximate the PDOS with a Debye model (a Debye model assumes linear dispersions, *i.e.*, phonon energy is a linear function of its momentum).

$$g(E) = 3E^2/E_D^3; \quad E \leq E_D$$
$$g(E) = 0; \quad E > E_D$$
, (8)

where $E_D$ is the Debye energy cutoff. In this framework we have,

285  $$\langle F \rangle = \frac{M}{\hbar^2} \int_0^{E_D} g(E) E^2 dE = \frac{3}{5} \frac{M}{\hbar^2} E_D^2, \quad (9)$$

286  so,

287  $$\int_0^{E_D} g(E) E^4 dE = \frac{3}{7} E_D^4 = \frac{25}{21} \frac{\hbar^4}{M^2} \langle F \rangle^2. \quad (10)$$

288  We can therefore rewrite the second term of Eq. 1 in the framework of a Debye model as,

289  $$1{,}000 \ln \beta_{I/I^*} = 1{,}000 \left( \frac{\hbar^2}{8k^2} \frac{\langle F \rangle}{T^2} - \frac{5\hbar^4}{2{,}016 k^4 M} \frac{\langle F \rangle^2}{T^4} \right) \left( \frac{1}{M^*} - \frac{1}{M} \right). \quad (11)$$

290  Replacing the constants with the relevant numerical values, we have for iron isotopic ratio
291  $^{56}$Fe/$^{54}$Fe, 1000 ln $\beta$ =2,904<F>/$T^2$-37,538<F>$^2$/$T^4$, with <F> in N/m and $T$ in K. This formula,
292  which is only valid for a Debye solid, can be generalized for any PDOS by writing,

293  $$1{,}000 \times \ln \beta = B_1 \frac{\langle F \rangle}{T^2} - B_2 \frac{\langle F \rangle^2}{T^4} \quad (12)$$

294  where $B_1$=2,904 and $B_2$ is a numerical value that depends on the shape of $g(E)$ or $S(E)$. Values of
295  $B_2$ were evaluated for 78 phases ranging from metals with Debye-like behaviors to hydroxides or
296  high-pressure phases with extended PDOS profiles. The exact values of 1000ln$\beta$ were calculated
297  in the temperature range 0-500 °C using Eq. B6 when only the PDOS was available or using Eq.
298  3 when raw NRIXS data were available. The first term (Eq. 11; 2,904<F>/$T^2$) was then
299  subtracted and the residuals were fitted with a function in 1/$T^4$. The values of $B_2$ range from
300  ~40,000 to 80,000 and depend on the nature of the phase (Fig. 4; $B_2 \approx$39,000 for biomolecules,
301  36,000 for metals, 51,000 for oxides and hydroxides, 64,000 for sulfates, 52,000 for sulfides,
302  68,000 for silicates, 80,000 for one carbonate). To a good approximation one can take $B_2$
303  constant and write,

304  $$1{,}000 \times \ln \beta = 2{,}904 \frac{\langle F \rangle}{T^2} - 52{,}000 \frac{\langle F \rangle^2}{T^4}, \quad (13)$$

305  with <F> in N/m and $T$ in K. The relative correction introduced by the second term is shown in
306  Fig. 3. For minerals relevant to low temperature geochemistry, using a constant $B_2$ value of
307  52,000 rather than measured ones introduces an inaccuracy of less than ~0.2 ‰ at 22 °C (Table
308  1), which is smaller than the overall uncertainty of the method. The approach outlined here is
309  particularly useful for calculating β-factors from published data as the average force constant is
310  often reported. Another significant advantage of the proposed method is that error propagation is
311  straightforward if the uncertainty on the force constant and $B_2$ are known and not correlated,

$$\sigma^2_{1000\times \ln\beta} = \left(\frac{B_1}{T^2} - 2B_2\frac{\langle F\rangle}{T^4}\right)^2 \sigma^2_{\langle F\rangle} + \left(\frac{\langle F\rangle^2}{T^4}\right)^2 \sigma^2_{B_2}. \qquad (14)$$

312  This approach is easier to implement than the one used by Polyakov and coworkers to treat
313  published NRIXS data. For example, Polyakov and Soultanov (2011) used previously published
314  NRIXS data (KOBAYASHI et al., 2007) to calculate the β-factor of chalcopyrite and obtained
315  values of 5.4 and 0.81 ‰ for the $^{56}Fe/^{54}Fe$ ratio at 22 and 500 °C, respectively. Kobayashi et al.
316  (2007) reported an average force constant of 146±2 N/m for this mineral. Based on Eq. 12, we
317  calculate for the $^{56}Fe/^{54}Fe$ ratio, 1,000×ln β=4.71 and 0.71 ‰ at 22 and 500 °C, respectively. The
318  discrepancy between Polyakov and Soultanov (2011) and the present study (5.4-4.7=0.7 ‰)
319  reflects a difference in the force constant estimate as it persists at high temperature, where the
320  second terms in Eqs 1 and 3 becomes negligible. The reason for this discrepancy is unknown but
321  it could reflect a problem of background subtraction. Kobayashi et al. (2007) acquired NRIXS
322  data over a symmetric energy interval (form -80 to +80 meV) and calculated the force constant
323  from the third order moment of $S(E)$, so the force constant that they report should be relatively
324  insensitive to background subtraction, as long as the background is constant. On the other hand,
325  Polyakov and Soultanov (2011) used the PDOS $g(E)$, which is sensitive to the assumed
326  background and can lead to erroneous results.

327  In Table 1, we summarize our estimates of β-factors for several molecules and minerals
328  based on published force constants and Eq. 12. For several of these phases, including
329  deoxymyoglobin, metmyoglobin, cytochrome f (oxidized and reduced), and orthoenstatite, it is
330  the first time that β-factors are calculated. We warn the reader that previous NRIXS studies were
331  not done for the purpose of estimating force constants and some equilibrium fractionation factors
332  may need to be reevaluated when better quality measurements become available.

333  There is overall good agreement between the results presented here and those reported by
334  Polyakov and coworkers (Table 1; Fig. 5). The discrepancies concern mostly high-pressure
335  phases with strong bonds. For example, the β–factors calculated for ferropericlase by Polyakov
336  (2009) were too low, sometimes by 30 % (*e.g.*, 2.1 *vs.* 3.2 ‰ at 109 GPa). The reason for this
337  discrepancy is unknown but it could be a smoothing or digitization artifact. In general, one
338  should be particularly cautious with high-pressure phases as they have extended high-energy tails
339  and as such, they are particularly prone to biases.

340

## 4. Results

The NRIXS spectra of $^{57}$Fe-enriched goethite, potassium-jarosite, and hydronium jarosite were measured. Figure 6 shows these spectra after removal of the elastic peak following the prescriptions of STURHAHN (2004). The spectra have been normalized, so the integral $\int_{E_{\min}}^{E_{\max}} S(E)\,\mathrm{d}E$ is equal to the recoil fraction during resonant nuclear scattering (1-$f_{lm}$, where $f_{lm}$ is the Lamb-Mössbauer factor). Table 2 summarizes all the thermo-elastic properties that can be retrieved from such spectrum, namely the Lamb-Mössbauer factor, the mean kinetic energy/atom, the mean force constant, the Lamb-Mössbauer factor at T=0 K, the mean kinetic energy at T=0 K, the vibrational specific heat, the vibrational entropy, and the critical temperature. For Fe isotope geochemistry, the critical number is the mean force constant <F>. As discussed in Sect. 3, this value is obtainable either directly from the NRIXS data or indirectly from the PDOS (Eq. 6). The force constants obtained using NRIXS data *S(E)* (the preferred method) are 314.10±9.66, 264.48±5.87, and 309.71±9.24 N/m for goethite, potassium-jarosite, and hydronium-jarosite, respectively. The force constants obtained using the PDOS *g(E)* are 316.57±10.83, 270.83±7.04, and 307.26.7±12.67 N/m for goethite, potassium-jarosite, and hydronium-jarosite, respectively.

While visual inspection of the NRIXS spectra suggests that there is no signal beyond -70 to +80 meV, the force constant is calculated by integrating an integrand that multiplies *S(E)* by $E^3$, thus dramatically amplifying the contribution of the high-energy tails. The relevant quantity to assess whether the background has been reached is therefore $S(E) \times E^3$, which extends to lower and higher energies (middle vertical panels of Fig. 6). Ideally, the force constant integral should plateau as the integration limits broaden, meaning that no signal is present above background. The value of the force constant integral is plotted in the right panels of Fig. 6. In the case of potassium-jarosite, a plateau is clearly reached. Despite the wide energy scans performed in this study, goethite and hydronium-jarosite do not show well-defined plateaus. As discussed in Sect. 3, the PHOENIX software can correct for truncated multiple phonon contributions by extrapolating the raw data beyond the acquired energy range. Comparison of the force constants before and after refinement shows that the correction is small (306.9 vs. 314.1 N/m for goethite, 302.4 vs. 309.7 N/m for H-jarosite, and 262.0 vs. 264.5 N/m for K-jarosite). The data thus appear reliable. Note that this extrapolation method cannot account for the presence of normal vibration modes outside of the scanned energy range and those buried under the background. Raman/IR and theoretical calculations can provide clues on the energy range required to capture all normal vibration frequencies. If present, high-energy modes could be below background, which would make them difficult to detect by NRIXS even if the energy range is expanded.

The same analysis was performed on the determination of the force constant from the PDOS (Fig. 7). At first sight, the PDOS of all three phases have reached background values at ~100 meV. However, the relevant integrand for calculation of the force constant is $g(E) \times E^2$, shown in the middle vertical panels of Fig. 7. For potassium-jarosite, no significant signal is present above ~100 meV, which is confirmed by examination of the force constant integral (right panels of Fig. 7). However, high-energy phonons seem to be present in hydronium-jarosite and goethite, all the way to 110 meV and possibly higher.

Background subtraction can affect the force constant derived from the PDOS and, to a lesser degree, that derived from $S(E)$. The background counts per channel were estimated by averaging the counts measured in the lowest 10 meV of the scans (*i.e.*, from -120 to -110 meV) and the 2σ uncertainty on the average was calculated. The background was then tuned within the limits allowed by the uncertainties to minimize the detailed balance (a balance between phonon annihilation and creation that depends on temperature). As expected, the force constant derived from the PDOS $g(E)$ is more sensitive to the background than that derived from $S(E)$.

In order to take into account the possibility that (1) some NRIXS signal may still be present at 130 meV and (2) background subtraction may not be optimum for the high-energy tail of the PDOS, we have added quadratically ±10 N/m of systematic error to the statistical uncertainty given by the PHOENIX software. This was assessed by testing the sensitivity of force constant estimates to background subtraction and refinement for missing multiple phonon contribution. The resulting equilibrium fractionation of the three minerals investigated for the $^{56}Fe/^{54}Fe$ ratio are given below (T is in K),

    Goethite:      $1{,}000 \times \ln \beta = 9.12 \times 10^5/T^2 - 6.11 \times 10^9/T^4$

    K-Jarosite:    $1{,}000 \times \ln \beta = 7.68 \times 10^5/T^2 - 4.08 \times 10^9/T^4$

    H$_3$O-Jarosite: $1{,}000 \times \ln \beta = 9.00 \times 10^5/T^2 - 6.23 \times 10^9/T^4$

To obtain the β-factors for the $^{57}Fe/^{54}Fe$ ratio, one can multiply the factors on the right by ~1.475, corresponding to a high-temperature approximation (MATSUHISA et al., 1978; YOUNG et al., 2002). The β-factors of K-Jarosite and H$_3$O-Jarosite are the first ever published and no direct comparison can be made with previous studies. Using previously published Mössbauer data for goethite (De Grave and Van Alboom 1991), Polyakov and Mineev (2000) calculated at 25 °C, $1,000 \times \ln\beta = +5.9$ ‰ while the NRIXS data presented here give an isotopic fractionation factor of ~$+9.49 \pm 0.38$ ‰. The β-factor derived from Mössbauer data uses the second-order Doppler shift, which is difficult to determine experimentally. As discussed in the introduction and by Polyakov et al. (2007), NRIXS β-factors are more accurate and should supersede Mössbauer data whenever available. For example, Polyakov and Mineev (2000) calculated a β-value for magnetite at 25 °C of +10.8 ‰ based on conventional Mössbauer data (De Grave et al. 1993), which was subsequently revised by Polyakov et al. (2007) to +7.3 ‰ based on new NRIXS data (SETO et al., 2003) and new conventional Mössbauer data. We calculate a value of +7.14 ‰ for magnetite (Table 1). The NRIXS-value is in good agreement with the experimentally measured equilibrium isotopic fractionation between fayalite and magnetite (SHAHAR et al., 2008). For hematite, Mössbauer gives $1,000 \times \ln \beta = +7.79$ ‰ for the $^{56}Fe/^{54}Fe$ ratio at 25 °C (POLYAKOV et al., 2007) while the NRIXS data of Sturhahn et al. (1999) give a value of $+7.33 \pm 0.27$ ‰ (Table 1). In this case, there is excellent agreement between the two approaches. A DFT-based quantum chemical calculation gives a value of 7.1 ‰ for hematite (Blanchard et al. 2009).

## 5. Discussion

*Goethite*

In the past decade, iron oxyhydroxides have been the focus of extensive iron isotope studies (DAUPHAS and ROUXEL, 2006; JOHNSON et al., 2008). Iron isotope measurements of ferric iron oxides have been used to understand subsurface fluid transport (BUSIGNY and DAUPHAS, 2007; CHAN et al., 2006; TEUTSCH et al., 2005), the conditions prevailing during banded-iron formation (CRADDOCK and DAUPHAS, 2011; DAUPHAS et al., 2004; HEIMANN et al., 2010; JOHNSON et al., 2003), the nature of metamorphosed rocks and the pressure-temperature conditions that affected them (DAUPHAS et al., 2007a; DAUPHAS et al., 2007b; FROST et al., 2007). Goethite was detected on Mars in the Columbia Hills at Gusev Crater, providing evidence of aqueous activity on the planet (MORRIS et al., 2006, 2008). To properly interpret Fe isotope variations in the rock record, one must know equilibrium fractionation factors. Several laboratory experiments have studied isotopic fractionation in systems involving goethite. Crosby et al. (2005, 2007a) studied experimentally the isotopic fractionation associated with biological dissimilatory iron reduction (DIR) of goethite, a form of microbial respiration that uses ferric iron as electron acceptor. Icopini et al. (2004) and Jang et al. (2008) examined Fe isotopic fractionation between Fe(II)$_{aq}$ and goethite. They argued that isotopic fractionation during sorption of Fe(II)$_{aq}$ onto the solid goethite substrate was the main factor controlling iron isotopic fractionation in their experiments. Mikutta et al. (2009) carried out an experiment where Fe(II)$_{aq}$ flowed through goethite-coated sandstone and they found that significant isotopic fractionation was present between Fe(II)$_{aq}$ and adsorbed Fe(II). All these experiments illustrate how complex fluid-rock interactions can be in the presence of a highly reactive mineral such as goethite.

Beard et al. (2010) recently estimated the equilibrium Fe isotopic fractionation between Fe(II)$_{aq}$ and goethite. They used a three-isotope technique first pioneered for O isotope geochemistry by Matsuhisa et al. (1978). Shahar et al. (2008) used the same technique to measure Fe isotopic fractionation between magnetite and fayalite. This approach uses an isotope spike (*e.g.*, $^{57}$Fe) to track the extent of equilibration between the reactants (at equilibrium, the reactants should plot on a single mass fractionation line). However, application of this technique does not guarantee accuracy if equilibrium has not been reached and kinetic processes such as diffusion or dissolution-re-precipitation reactions control isotopic exchange between the reactants, or if more than two chemical species are involved in the reaction. Beard et al. used a stepwise dissolution protocol to separate the various iron components Fe(II)$_{aq}$, sorbed Fe(II), surface Fe(III), and bulk goethite. They obtained an equilibrium isotopic fractionation between Fe(II)$_{aq}$ and goethite of -1.05±0.08 ‰ at 22 °C for the $^{56}$Fe/$^{54}$Fe ratio. Comparing this value with the results presented here is not straightforward as Beard et al. measured the difference 1,000×ln $\beta_{Fe(II)aq}$-1,000×ln $\beta_{goethite}$, while we calculated 1,000×ln $\beta_{goethite}$. Therefore, one needs to know 1,000×ln $\beta_{Fe(II)aq}$ to compare the two studies.

Several studies based on electronic structure theory or other computational methods have reported the values of $\beta_{Fe(II)aq}$ and/or $\beta_{Fe(III)aq}$ (ANBAR et al., 2005; DOMAGAL-GOLDMAN and KUBICKI, 2008; HILL and SCHAUBLE, 2008; HILL et al., 2009; HILL et al., 2010; OTTONELLO and ZUCCOLINI, 2009; RUSTAD et al., 2010; RUSTAD and DIXON, 2009; SCHAUBLE et al., 2001). These calculations match experimentally measured values for the difference $\beta_{Fe(III)aq}$-$\beta_{Fe(II)aq}$ (HILL et al., 2009; WELCH et al., 2003). However, it remains uncertain at the present time whether the β-factors are accurate on an absolute scale. Anbar et al. (2005) reported β-values for the $^{56}$Fe/$^{54}$Fe ratio of Fe(II)$_{aq}$ at 25 °C from 6.23 to 6.69 ‰; Ottonello and Zuccolini (2009) reported values from 4.859 to 5.537 ‰; Rustad et al. (2010) reported a value of ~4.67 ‰;

Domagal-Goldman and Kubicki (2008) reported values of ~5.66-6.55 ‰; Hill and Schauble (2008) reported values of 5.42 to 6.70 ‰. Overall, theoretical calculations give a β-value for $Fe(II)_{aq}$ that ranges over 2 ‰ from 4.6 to 6.7 ‰. Combining these predictions with the β-factor that we calculated for goethite (9.67±0.39 ‰ at 22 °C and 9.49±0.38 ‰ at 25 °C), we would predict an equilibrium isotopic fractionation factor between goethite and $Fe(II)_{aq}$ of +3 to +5 ‰; most of the uncertainty stems from the large variability in predictions from electronic structure theory. This is much larger than the equilibrium fractionation of +1.05 ‰ that was estimated by Beard et al. Clearly, some of these estimates must be in error.

Polyakov and Soultanov (2011) recently reviewed all existing constraints on mineral-fluid equilibrium isotopic fractionation and proposed that the β value of $Fe(II)_{aq}$ at 25 °C be revised to 4.9 ‰, in agreement with some *ab-initio* calculations, in particular that of Rustad et al. (2010) who explicitly represented the second water solvation shell. Equilibrium isotopic fractionation between $Fe(II)_{aq}$ and $Fe(III)_{aq}$ was measured for several ferric-iron bearing minerals. Wu et al. (2011) determined a fractionation of +3.17±0.08 ‰ between hydrous ferric oxide (ferrihydrite) and $Fe(II)_{aq}$. Skulan et al. (2002) found no measurable equilibrium isotopic fractionation between hematite and $Fe(III)_{aq}$ (*i.e.*, +0.1±0.2 ‰ at 98 °C, corresponding to +0.15±0.30 at 25 °C assuming that 1000 ln β scales as $1/T^2$). Given that the equilibrium Fe isotopic fractionation between $Fe(III)_{aq}$ and $Fe(II)_{aq}$ is +2.9±0.5 ‰ at 25 °C (WELCH et al., 2003), the equilibrium isotopic fractionation between hematite and $Fe(II)_{aq}$ is +3.0±0.6 ‰. To summarize, the equilibrium isotopic fractionations for both ferrihydrite-$Fe(II)_{aq}$ and hematite-$Fe(II)_{aq}$ are ~+3 ‰, while Beard et al.'s experiment suggest that the isotopic fractionation goethite-$Fe(II)_{aq}$ is +1 ‰. A +2 ‰ equilibrium isotopic fractionation between hematite and goethite at 25 °C would require a force constant for goethite of ~180 N/m (assuming that the force constant of hematite is 244 N/m; Table 1). This is similar to values measured in $Fe^0$ or $Fe^{2+}$-bearing phases and is very unlikely given that most $Fe^{3+}$-bearing phases tend to have high force constant (*i.e.*, 244 N/m for hematite, 264 N/m for potassium-jarosite, 310 N/m for hydronium-jarosite, and 314 N/m for goethite; Fig. 8). The higher force constants for hydronium-jarosite and goethite relative to other $Fe^{3+}$-bearing phases might be explained if $OH^-$ forms stiffer Fe-O bonds than are found in anhydrous minerals.

Our results call for a reexamination of exchange experiments between goethite and $Fe(II)_{aq}$. One way to assess whether the measured isotopic fractionation reflects equilibrium would be to carry out experiments at higher temperature as equilibrium isotopic fractionation should decrease as $1/T^2$ while kinetic isotope effects may still be present at higher temperature (*i.e.*, one can measure the temperature dependence of the fractionation factor and test whether it conforms to expectations for equilibrium). We note that Crosby et al. (2007) had measured a fractionation between Fe(II)$_{aq}$ and surface Fe(III) in goethite of +2.62±0.57 ‰, which would be in better agreement with the value documented here.

*Jarosite*

The Mars rovers have revealed the occurrence of iron-rich rocks composed of hematite spherules embedded in a jarosite matrix (KLINGELHÖFER et al., 2004). This detection has provided unambiguous evidence for the presence of liquid water at the surface of Mars in the past. Considerable uncertainty remains, however, as to the chemical pathways that permitted deposition of rocks composed of such an unusual mineral assemblage. Interpretation is complicated by the fact that these rocks have no direct terrestrial analogues available for comparison. The field occurrence of hematite spherules is reminiscent of the iron-oxide plus quartz concretions found in the Navajo sandstone in Utah, yet the sedimentology and details of the composition of the concretions of the latter formation has no similarity with the martian rocks (BUSIGNY and DAUPHAS, 2007; CHAN et al., 2004; Morris et al. 2005). Some rocks near fumaroles in Hawaii contain jarosite and hematite spherules, representing the best terrestrial analogues of the martian outcrops (MORRIS et al., 2005). There are however major differences between these rocks and the martian samples, starting with the size of the spherules. Iron isotopes have been used on Earth to understand how associated jarosite-hematite mineral assemblages were formed and to trace Fe mobilization during acid sulfate alteration (DAUPHAS and MORRIS, 2008). When returned samples are available, a similar approach can be used to study martian surface geology. Such Fe isotope measurements will allow us to address the origin of Fe in hematite-jarosite and the chemical pathways that led to the formation of these phases. It will be possible to address two questions:

1. What is the provenance of the chemical sedimentary component in these formations? It was suggested that the source of iron involved acid sulfate alteration of basalts (MCLENNAN et al., 2005; MORRIS et al., 2005; TOSCA et al., 2005). Studies of natural samples and laboratory experiments have shown that such alteration processes would produce fluids that have light Fe isotopic composition relative to the substrate (BRANTLEY et al., 2004; ROUXEL et al., 2003). The Fe isotopic composition of martian basaltic rocks is known from measurements of SNC meteorites, which average a $\delta^{56}$Fe value of ~+0.02 ‰ relative to reference material IRMM-014 (ANAND et al., 2006; POITRASSON et al., 2004; WEYER et al., 2005; Wang et al., 2012). We thus expect the alteration fluid to have a $\delta^{56}$Fe value of ~-0.2 ‰. Our NRIXS results show that at equilibrium, potassium-jarosite should have a $\delta^{56}$Fe value higher than Fe(III)$_{aq}$ by ~+1 ‰ (Table 1 for K-jarosite; POLYAKOV and SOULTANOV, 2011 for Fe(III)$_{aq}$). The isotopic composition of jarosite will also depend on the extent of the conversion of Fe(III)$_{aq}$ into jarosite. The fluid-rock ratio may have been limited and water-rock interactions may have been short-lived. In this framework, one would expect jarosite to have a $\delta^{56}$Fe value close or slightly higher than the fluid.

2. Various geochemical pathways are possible for the formation of hematite spherules. Hematite spherules and jarosite can form by direct precipitation under hydrothermal conditions (Golden et al. 2008). Other scenarios involve conversion of jarosite into goethite or hematite upon fresh water recharge (TOSCA et al., 2005; ZOLOTOV and SHOCK, 2005), or oxidation of a late-stage evaporite mineral such as melanterite (TOSCA et al., 2005). If the conversion is quantitative and is localized, one would expect hematite to inherit the Fe isotopic composition of the precursor mineral. On the other hand, the transformation may reflect a displacement of equilibrium conditions in the whole system. Measuring the Fe isotopic composition of hematite and jarosite will provide a means to assess the degree of disequlibrium between phases. If hematite has the exact same $\delta^{56}$Fe value as jarosite, this will support the view that hematite was produced by destabilization of jarosite during a recharge event.

Much of this discussion is speculative but it illustrates the powerful constraints that Fe isotopes can provide to interpret the record of iron-rich sediments observed on the martian surface. The fractionation factors calculated here for goethite, hematite, and jarosite will provide a framework to interpret Fe isotope variations in returned martian samples. Sooner, they will help interpret Fe isotope variations documented in terrestrial analogues of martian sediments and they will help understand iron-transport in acid mine drainage.

## 6. Conclusion

A new approach to determine equilibrium fractionation factor from NRIXS data is presented. It relies on the fact that the reduced partition function ratio is related through a Bernoulli expansion to the moments of the NRIXS spectrum $S(E)$ (Eq. 3; Appendix C). At high temperature, $1000\ln\beta = B_1 <F>/T^2$, with $B_1 = 2,904$ and $<F>$ the average force constant of the bonds involving Fe atoms (Eq. 7), which is easily obtained by taking the third-order moment of the NRIXS spectrum (Eq. 6).

The force constants and the equilibrium fractionation factors that are derived are highly sensitive to the high-energy tails of the NRIXS spectrum. Such measurements are therefore highly demanding as they require very long acquisition times over broad energy intervals to accurately measure the tails of the nuclear resonance scattering spectrum. This difficulty may explain the discrepancy found for high-pressure phases between the β-factors derived from the force constant approach (this study) compared with those inferred by Polyakov (2009). For most low-pressure minerals, there is good agreement between our approach and the results reported by Polyakov et al. (2007), Polyakov (2009), and Polyakov & Soultanov (2011) based on the Thirring expansion of the kinetic energy in moments of the PDOS $g(E)$.

The approach that we propose based on the moments of $S(E)$ (Eq. 3, Appendix C) is more reliable than that based on the moments of $g(E)$ (Eq. 1; Polyakov et al. 2005b, 2007) as it is less sensitive to background subtraction, it allows quantification of the contribution to the β-factor of multiple phonons outside the acquisition energy window, and it yields smaller statistical uncertainties.

We have applied the force constant approach to published NRIXS data and report 1000 ln β for important phases of geological and biochemical relevance such as myoglobin, cytochrome f, orthoenstatite, metal, troilite, chalcopyrite, hematite, and magnetite (Table 1).

The NRIXS spectra of goethite, potassium-jarosite, and hydronium-jarosite are reported (Fig. 6). These are critical phases for aqueous, low temperature iron isotope geochemistry. The goethite fractionation factor derived from NRIXS (Fig. 8) disagrees with values obtained by combining laboratory exchange experiments (Beard et al. 2010) and calculations based on electronic structure theory (Rustad et al. 2010). The laboratory exchange experiments give β-factors that differ markedly between hematite and hydrous ferric oxide on the one hand (Skulan et al. 2002; Wu et al. 2011), and goethite on the other hand (Beard et al. 2010).

Further experimental work will be needed but we suspect that the equilibrium fractionation obtained in laboratory exchange experiments is too low for goethite.

Our work shows that the relationship between iron equilibrium isotopic fractionation factor and force constant provides a direct method of calculating β-factors in high-temperature/high-pressure systems using NRIXS data (Eq. 6, 7). High-pressure phases have extended NRIXS spectra, requiring wide energy scans to calculate iron force constants. Furthermore, the samples in diamond anvil cells are small, requiring long acquisition times. These two factors make the determination of iron force constants in high-pressure phases particularly challenging, so published data should be considered with a grain of salt for application to iron isotope geochemistry.

The formalism developed here will be extremely useful to analyze NRIXS data for application to the isotope geochemistry of elements with low lying nuclear excited states such as Dy, Eu, Kr, and Sn.

**Acknowledgements.** We thank C. Achilles for determination of particle sizes using XRD data. Discussions with W. Sturhahn, R.N. Clayton, R. Caracas, and T. Fujii regarding data reduction, background subtraction, and stable isotope fractionation were greatly appreciated. M. Meheut, V. Polyakov, A. Shahar, and Associate Editor E. Schauble are thanked for their thoughtful reviews of the manuscript. This work was supported by NASA (NNX09AG59G), by NSF EAR Petrology and Geochemistry (EAR-1144429), and by a Packard Fellowship to N. Dauphas. L. Gao acknowledges the financial support from COMPRES under NSF Cooperative Agreement EAR 10-43050. Use of the Advanced Photon Source, an Office of Science User Facility operated for the U.S. Department of Energy (DOE) Office of Science by Argonne National Laboratory, was supported by the US DOE under contract N° DE-AC02-06CH11367.

**Appendix A.**

Mathematica data reduction script (available online at http://...).

**Appendix B.**

Below, we demonstrate how equilibrium iron isotope fractionation factors at any temperature can be derived from the moments of the partial phonon density of states g(E). Bigeleisen (1958) carried out a Bernoulli expansion of the reduced partition function ratio and obtained a formula that is valid for dimensionless frequencies $\hbar\omega_{max}/kT < 2\pi$ (also see Elcombe and Hulston 1975).,

$$\ln \beta_{I/I^*} = \frac{1}{\widetilde{N}} \sum_{j=2}^{\infty} \sum_{i}^{3N} \frac{B_j \delta u_i^j}{j! \, j}, \qquad (B1)$$

where $\widetilde{N}$ is the number of isotopically substituted atoms, $N$ is the total number of atoms in a multi-atomic compound, $\delta u_i^j = u_i^{*j} - u_i^j$ ($u^*$ and $u$ are the dimensionless frequencies of $i^{th}$ phonon mode for the two substituted iron isotopes; $u_i = \hbar\omega_i/kT$), and $B_j$ are the Bernoulli numbers (*i.e.*, $B_0=1$, $B_1=-1/2$, $B_2=1/6$, $B_3=0$, $B_4=-1/30$, $B_5=0$, $B_6=1/42$).

Elcombe (1974) and Menéndez et al. (1994) demonstrated using first-order perturbation theory that normal mode frequency shifts upon isotope substitution are related to the masses of the substituted isotopes,

$$\frac{u_i^{*2} - u_i^2}{u_i^2} = \frac{M - M^*}{M^*} \cdot \frac{1}{N} \sum_{v} |e_{iv}|^2, \qquad (B2)$$

where $v$ denotes the $\widetilde{N}$ isotopically substituted atoms. Only one isotope species is replaced, so the mass change is constant. $|e_{iv}|^2$ is the square modulus of the polarization vector for mode $i$ and isotope $v$ ($e_{iv}$ is a vector of dimension 3, corresponding to the 3 directions of space). The phonon polarization vectors are normalized to $N$. The polarization vector is a mathematical construction that relates the displacement of a particular atom to the excitation of a particular vibration mode. Given that $x = u^*/u$ is close to 1 and $x^j - 1 = (1 + x - 1)^j - 1 \simeq j(x - 1)$, it follows,

$$\frac{u_i^{*j} - u_i^j}{u_i^j} \approx j\left(\frac{u_i^* - u_i}{u_i}\right) \approx \frac{j}{2}\left(\frac{M - M^*}{M^*}\right) \cdot \frac{1}{N} \sum_{v} |e_{iv}|^2, \qquad (B3)$$

Using this formula, Eq. B1 can be rewritten as (this is Eq. 13 of Elcombe and Hulston 1975),

$$\ln\beta_{I/I^*} = \frac{1}{2}\left(\frac{M}{M^*}-1\right)\sum_{j=2}^{\infty}\frac{B_j}{j!}\frac{1}{\widetilde{N}}\sum_{v}\frac{1}{N}\sum_{i}^{3N}|e_{iv}|^2 u_i^j, \qquad (B4)$$

where the last two summations over all substituted isotopes and all phonon modes can be expressed in terms of the moments of the projected partial phonon DOS (a definition is provided in Hu et al. 2003; Eq. 2),

$$\ln\beta_{I/I^*} = \frac{1}{2}\left(\frac{M}{M^*}-1\right)\sum_{j=2}^{\infty}\frac{B_j}{j!}3\frac{m_j^g}{(kT)^j} = \frac{3}{2}\left(\frac{M}{M^*}-1\right)\sum_{j=2}^{\infty}\frac{B_j}{j!\,(kT)^j}m_j^g, \qquad (B5)$$

with $m_j^g = \int_0^{+\infty} E^j g(E)dE$ the $j^{th}$ moment of iron partial PDOS $g(E)$ (note that the integral of the PDOS is $\int g(E)dE = 1$). Here the sample is assumed to be isotropic or powder. The factor of 3 comes from comparing the square modulus of the phonon polarization vector *vs.* its projection along a particular direction. We recognize the sum $\sum_{j=0}^{\infty} B_j x^j/j! = x/(e^x-1)$, so Eq. B5 equation can be rewritten as,

$$\ln\beta_{I/I^*} = \frac{3}{2}\left(\frac{M}{M^*}-1\right)\int_0^{E_{max}}\left(\frac{E}{2kT}+\frac{E/kT}{e^{E/kT}-1}-1\right)g(E)dE. \qquad (B6)$$

Equations B5 and B6 had been derived by Polyakov at al. (2005b, 2007) using a different approach based on the kinetic energy. Note that there is an error in Eq. 9 of Polyakov et al (2005b); the factor should be $h/k$ rather than $(h/k)^{2i}$.

**Appendix C.**

The derivation below establishes a relationship between the moments of $g$ and the moments of $S$.

Let us denote,

$m_{2n+1}^{S_1} = \int_{-\infty}^{+\infty} E^{2n+1} S_1(E) dE$ the 2n+1 moment of $S_1(E)$,

$m_{2n}^g = \int_0^{+\infty} E^{2n} g(E) dE$ the 2n moment of $g(E)$.

In the harmonic approximation, the one phonon term in the excitation probability is related to $g(|E|)$ through,

$$S_1(E) = \frac{f E_R g(|E|)}{E\left(1-e^{-E/kT}\right)}, \qquad (C1)$$

where $f$ is the Lamb-Mössbauer factor. The 2n+1 (odd) moment of $S_1(E)$ with $n$ an integer is,

659 $$m^{S_1}_{2n+1} = \int_{-\infty}^{0} E^{2n+1} S_1(E) dE + \int_{0}^{+\infty} E^{2n+1} S_1(E) dE. \quad (C2)$$

660 Using Eq. C1, the first integral in Eq. C2 can be rewritten as,

661 $$\int_{-\infty}^{0} E^{2n+1} S_1(E) dE = \int_{0}^{+\infty} E^{2n+1} \frac{f E_R g(E)}{E(1-e^{E/kT})} g(E) dE, \quad (C3)$$

662 while the second integral takes the form,

663 $$\int_{0}^{+\infty} E^{2n+1} S_1(E) dE = \int_{0}^{+\infty} E^{2n+1} \frac{f E_R g(E)}{E(1-e^{-E/kT})} g(E) dE. \quad (C4)$$

664 We therefore have (C3+C4),

665 $$m^{S_1}_{2n+1} = f E_R \int_{0}^{+\infty} \frac{E^{2n+1}(1-e^{E/kT}) + E^{2n+1}(1-e^{-E/kT})}{E(2-e^{-E/kT} - e^{E/kT})} g(E) dE. \quad (C5)$$

666 It follows,

667 $$m^{S_1}_{2n+1} = f E_R \int_{0}^{+\infty} E^{2n} g(E) dE = f E_R m^{g}_{2n}. \quad (C6)$$

668 For a *2n* (even) moment, the temperature terms do not cancel out and we have instead,

669 $$m^{S_1}_{2n} = f E_R \int_{0}^{+\infty} E^{2n-1} \left( \frac{1}{1-e^{-E/kT}} - \frac{1}{1-e^{E/kT}} \right) g(E) dE \quad (C7)$$

670

671 For energies that are most relevant to higher moment calculations in phases with high force constants, the term in parenthesis is approximately 1,

672 $$m^{S_1}_{2n} \approx f E_R m^{g}_{2n-1}. \quad (C8)$$

673

674 This is the same formula as the one that we had derived for odd moments of *S* (C6). This approximation can be refined by recognizing that,

675 $$\frac{1}{1-e^{-E/kT}} - \frac{1}{1-e^{E/kT}} \approx a + \frac{b}{E/kT}. \quad (C9)$$

676

The figure below shows these two equations at T=300 K with a=0.63 and b=1.48.

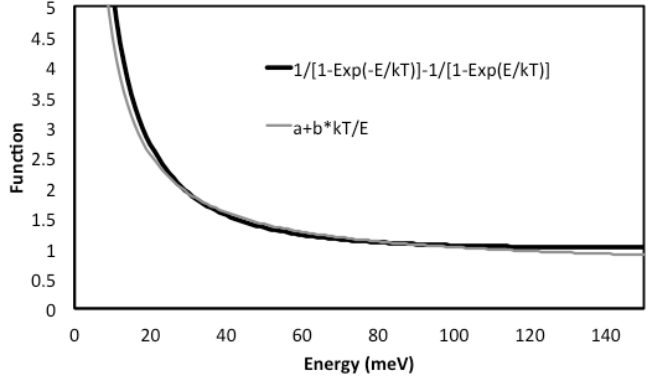

677
678

The even moments then take the form,

679 $m_{2n}^{S_1} \approx afE_R m_{2n-1}^{g} + bkTfE_R m_{2n-2}^{g}$.  (C10)

680

Thus, the moments of $S_1$ can be related to the moments of $g$. The moments introduced here are
681
centered on zero. Let us denote,

682 $R_i^{S_n} = \int_{-\infty}^{+\infty} (E - E_r)^i S_n(E) dE$  (C11)

683
the moment of the n-phonon excitation probability centered on $E_r$. Because $E_r \ll E$ over most of
684
the PDOS,

685 $R_i^{S_n} \approx \int_{-\infty}^{+\infty} (E^i - iE^{i-1} E_r) S_n(E) dE = m_i^{S_n} - iE_r m_{i-1}^{S_n}$.  (C12)

686

We are now going to establish a relation between the moments of the *n*-phonons and the
687
moments of the 1 phonon. In the harmonic approximation, the *n*-phonon is calculated using,

$$S_n(E) = \frac{1}{nf} \int_{-\infty}^{+\infty} S_{n-1}(x) S_1(E - x)\, dx. \quad (C13)$$

688
The *i*-th moment centered on zero is then,

689 $m_i^{S_n} = \frac{1}{nf} \int_{-\infty}^{+\infty} \int_{-\infty}^{+\infty} S_{n-1}(x) S_1(E-x) E^i \, dx dE$.  (C14)

690

permuting the integration order and changing the variable to $u=E-x$, this equation takes the form,

691 $m_i^{S_n} = \frac{1}{nf} \int_{-\infty}^{+\infty} S_{n-1}(x) \int_{-\infty}^{+\infty} S_1(u)(u+x)^i \, du dx$.  (C15)

692

We can now do a binomial expansion of $(u+x)^i$,

693 $m_i^{S_n} = \frac{1}{nf} \int_{-\infty}^{+\infty} S_{n-1}(x) \int_{-\infty}^{+\infty} S_1(u) \sum_{k=0}^{i} C_i^k u^{i-k} x^k \, du dx$,  (C16)

where $C_i^k = i!/[(i-k)!k!]$ are the binomial coefficients. Taking the summation out of the integral it follows,

$$m_i^{S_n} = \frac{1}{nf} \sum_{k=0}^{i} C_i^k \int_{-\infty}^{+\infty} S_{n-1}(x) x^k \int_{-\infty}^{+\infty} S_1(u) u^{i-k} \, du dx. \quad (C17)$$

We therefore have,

$$m_i^{S_n} = \frac{1}{nf} \sum_{k=0}^{i} C_i^k m_k^{S_{n-1}} m_{i-k}^{S_1}. \quad (C18)$$

This relationship allows one to calculate the moments of all multiple phonons using the moment of the single phonon. For example, the $i^{th}$-moment of the 2-phonon is,

$$m_i^{S_2} = \frac{1}{2f} \sum_{k=0}^{i} C_i^k m_k^{S_1} m_{i-k}^{S_1}. \quad (C19)$$

Knowing the moments of the 2-phonon, it is straightforward to calculate the moments of the 3-phonon and proceed iteratively to calculate the $i^{th}$ moment of any multiple phonon. The $i^{th}$ moment of $S(E)$, the excitation probability is the sum of all $i^{th}$ moments of $S_n(E)$,

$$m_i^S = m_i^{S_0} + m_i^{S_1} + \frac{1}{f} \sum_{n=2}^{\infty} \sum_{k=0}^{i} \frac{C_i^k m_k^{S_{n-1}} m_{i-k}^{S_1}}{n}, \quad (C20)$$

where $S_0$ is a delta function and its moment is zero for $i \geq 1$. With Eq. C6 and C9, we have shown that there is a relationship between $m_i^{S_1}$ and the moments of $g$. Equation C18 shows that there is a relationship between $m_i^{S_1}$ and $m_i^S$. We can thus express the moments of $S$ as a function of the moments of $g$ and vice versa. We have calculated the second to tenth moments of $S$ centered on $Er$ by using Eq. C11 and expanding Eq. C18 into moments of $S_1$, which were expressed as combinations of the moments of $g$ (Eq. C6 and C9). We used the fact that $m_0^g = 1$ and $m_0^{S_1} = -f \ln f$. Furthermore, the series

$$f - \ln f + \frac{1}{2} f (\ln f)^2 - \frac{1}{6} f (\ln f)^3 + \frac{1}{24} f (\ln f)^4 + \ldots \approx 1$$

appeared in many coefficients, simplifying the final expressions. The terms that were found to be negligible are omitted in the following equation,

$$R_2^S = aE_r m_1^g + bE_r kT_e$$

$$R_3^S = E_r m_2^g$$

$$R_4^S = aE_r m_3^g + 3a^2 E_r^2 \left(m_1^g\right)^2 + bE_r kT_e m_2^g + 6abE_r^2 kT_e m_1^g + 3b^2 E_r^2 k^2 T_e^2$$

$$R_5^S = E_r m_4^g + 10aE_r^2 m_1^g m_2^g + 10bE_r^2 kT_e m_2^g$$

$$R_6^S = aE_r m_5^g + 15a^2 E_r^2 m_1^g m_3^g + 10 E_r^2 \left(m_2^g\right)^2 + 15b^2 E_r^2 k^2 T_e^2 m_2^g + 15abE_r^2 kT_e m_3^g + 15abE_r^2 kT_e m_1^g m_2^g + bE_r kT_e m_4^g + 45a^2 bE_r^3 kT \left(m_1^g\right)^2$$

$$+ 45ab^2 E_r^3 k^2 T^2 m_1^g + 15b^3 E_r^3 k^3 T^3$$

$$R_7^S = E_r m_6^g + 35aE_r^2 m_2^g m_3^g + 21aE_r^2 m_1^g m_4^g + 105a^2 E_r^3 \left(m_1^g\right)^2 m_2^g + 35bE_r^2 kT_e \left(m_2^g\right)^2 + 21bE_r^2 kT_e m_4^g + 210abE_r^3 kT_e m_1^g m_2^g + 105b^2 E_r^3 k^2 T_e^2 m_2^g$$

(C21)

where $T_e$ denotes the temperature at which the experiments were performed (typically ~300 K). The second equation corresponds to Lipkin's sum rule for the force constant. To our knowledge, this is the first time that this formula is established on mathematical grounds alone. There are six unknowns (the moments of $g$) in 6 equations and this system can be solved,

$$m_2^g = \frac{R_3^S}{E_r}$$

$$m_4^g = \frac{R_5^S - 10 R_2^S R_3^S}{E_r}$$

$$m_6^g = \frac{R_7^S + 210\left(R_2^S\right)^2 R_3^S - 35 R_3^S R_4^S - 21 R_2^S R_5^S}{E_r}$$

(C22)

Although the derivations were made in the most general case (for the even moment of $S_l$, Eq. C10 was used rather than Eq. C8 and the relationship in Eq. C21 has some temperature dependence), the formulas that relate the even moments of $g$ to the moments of $S$ are temperature independent and should be valid for any phase.

**Figure captions**

**Fig. 1.** Experimental set-up used for NRIXS spectroscopy (modified from STURHAHN and JACKSON, 2007). The pulsed X-ray beam passes through two monochromators to achieve an energy dispersion of ±1 meV at 14.4 keV. This energy can induce a nuclear transition to an excited state of $^{57}$Fe, which decays to the ground state with a lifetime of 141 ns. The signal from electronic scattering is separated from that from nuclear scattering using the fact that the latter signal is produced with some delay. The energy spectrum is produced by scanning the energy (*e.g.*, -120 to +130 meV) with the high-resolution monochromator, and recording the cummulative signal from 15 to 140 ns after each pulse of the synchrotron beam.

**Fig. 2.** Comparison between the coefficients in the polynomial expansion 1,000ln $\beta=A_1/T^2+A_2/T^4+A_3/T^6$ obtained from *g(E)* (Eq. 1) and *S(E)* (Eq. 3, Appendix C). The samples were all measured at the APS using the same acquisition protocol as that used in the present study. There is good agreement between the various coefficients but those obtained from S(E) have uncertainties that are on average ~2 times (and up to 5 times) smaller than those obtained from *g(E)*. For this, and other reasons discussed in the text, Eq. 3 is preferred over Eq. 1.

**Fig. 3.** Relative departure from Eq. 7 introduced by the $B_2$ term in Eq. 17 for the $^{56}$Fe/$^{54}$Fe ratio. The relative departure is equal to $(B_1<F>/T^2-B_2<F>^2/T^4)/\ B_1<F>/T^2=1-(B_2/B_1)<F>/T^2$. With $B_1$=2,904 and $B_2\approx$52,000, the correction factor becomes 1-17.9$<F>/T^2$ (gray continuous lines). For high-temperature phases, to a good approximation we have 1000 ln $\beta\simeq$2,904$<F>/T^2$.

**Fig. 4.** Determination of $B_2$ used to calculate equilibrium isotopic fractionation factors ($^{56}$Fe/$^{54}$Fe) from the force constant; 1,000 ln $\beta$=2,904$<F>/T^2-B_2<F>^2/T^4$. $B_2$ is a parameter that relates the 4$^{th}$ moment of the PDOS to the 2$^{nd}$ moment. For a Debye solid, this value is 37,538. For other solids with more extended PDOS, the $B_2$ values can be higher and reach ~70,000. In most cases, adopting a constant value of 52,000 provides sufficient accuracy on calculated equilibrium isotopic fractionation factors as this is only a minor second order correction that affects phases with high $<F>$ values at low temperature.

764 **Fig. 5.** Comparison between 1,000 ln β ($^{56}$Fe/$^{54}$Fe) at 500 °C reported by Polyakov and
765 coworkers and values derived from force constants (Table 1). Overall, there is good agreement
766 between the two approaches except for a few high-pressure phases characterized by PDOS with
767 extended high-energy tails.

769 **Fig. 6.** NRIXS derivation of the average force constant for goethite (top panels), potassium-
770 jarosite (middle horizontal panels), and hydronium-jarosite (bottom panels). The force constant
771 is calculated from the NRIXS spectrum (left) by using the formula
772 $\langle F \rangle = \frac{M}{E_R \hbar^2} \int_{-\infty}^{+\infty} (E - E_R)^3 S(E) \mathrm{d}E$ (Eq. 6). While $S(E)$ reaches near-background values above 80
773 meV for all three phases, significant signal may still be present above that value, which gets
774 amplified in the integral when $S(E)$ is multiplied by the cube of the energy (middle vertical
775 panel). The right panels show the values of the force constant integrals for different values of the
776 integration limits; $\langle F \rangle = \frac{M}{E_R \hbar^2} \int_{-x}^{+x} (E - E_R)^3 S(E) \mathrm{d}E$ with x=0 to 130 meV. While the force constant
777 integrals never reach a plateau, the vertical middle panels show that little signal is expected
778 above 130 meV and the force constants that we derive are not affected by truncation in energy.
779 The values of the force constants used for calculation of β-factors are obtained from $S(E)$ after
780 refinement (see text for details), which takes into account the contribution of multiple phonons
781 outside of the acquisition range. The force constants derived from these data are 314.1±13.9,
782 264.5±11.6, and 309.7±13.6 N/m for goethite, potassium-jarosite, and hydronium-jarosite,
783 respectively (Table 2).

785 **Fig. 7.** PDOS derivation of the average force constant for goethite (top panels), potassium-
786 jarosite (middle horizontal panels), and hydronium-jarosite (bottom panels). The force constant
787 is calculated from the PDOS (left) by using the formula $\langle F \rangle = \frac{M}{\hbar^2} \int_0^{+\infty} g(E) E^2 \, \mathrm{d}E$ (Eq. 6). While the
788 PDOS reaches near-background values above 80 meV for all three phases, significant signal may
789 still be present above that value, which gets amplified in the integral when the PDOS is
790 multiplied by the square of the energy (middle vertical panel). The right panels show the values
791 of the force constant integrals for different values of the integration upper-limit; $\frac{M}{\hbar^2} \int_0^x g(E) E^2 \, \mathrm{d}E$
792 with x=0 to 130 meV. If the PDOS reaches background level, the integral should reach a plateau

793  corresponding to the average force constant of the chemical bond. While this is the case for the
794  jarosite samples, goethite never reaches a plateau. The vertical middle panels show that little
795  signal is expected above 130 meV and the force constants that we derive are not affected by
796  truncation in energy.
797
798

799  **Fig. 8.** 1,000 ln β ($^{56}$Fe/$^{54}$Fe) of $Fe^{3+}$-bearing minerals (goethite, jarosite, and hematite; Table 1),
800  $Fe^{3+}_{aq}$, and $Fe^{2+}_{aq}$ (POLYAKOV and SOULTANOV, 2011).
801
802
803
804

**References**

Adams, K. L., Tsoi, S., Yan, J., Durbin, S. M., Ramdas, A. K., Cramer, W. A., Sturhahn, W., Alp, E. E., and Schulz, C., 2006. Fe Vibrational Spectroscopy of Myoglobin and Cytochrome f. *The Journal of Physical Chemistry B* **110**, 530-536.

Alp, E., Sturhahn, W., Toellner, T., Zhao, J., Hu, M., and Brown, D., 2002. Vibrational Dynamics Studies by Nuclear Resonant Inelastic X-Ray Scattering. *Hyperfine Interactions* **144-145**, 3-20.

Alp, E. E. and et al., 2001. Lattice dynamics and inelastic nuclear resonant x-ray scattering. *Journal of Physics: Condensed Matter* **13**, 7645.

Anand, M., Russell, S. S., Blackhurst, R. L., and Grady, M. M., 2006. Searching for signatures of life on Mars: an Fe-isotope perspective. *Philosophical Transactions of the Royal Society B: Biological Sciences* **361**, 1715-1720.

Anbar, A. D., Jarzecki, A. A., and Spiro, T. G., 2005. Theoretical investigation of iron isotope fractionation between Fe(H2O)63+ and Fe(H2O)62+: Implications for iron stable isotope geochemistry. *Geochimica et Cosmochimica Acta* **69**, 825-837.

Beard, B. L., Handler, R. M., Scherer, M. M., Wu, L., Czaja, A. D., Heimann, A., and Johnson, C. M., 2010. Iron isotope fractionation between aqueous ferrous iron and goethite. *Earth and Planetary Science Letters* **295**, 241-250.

Bigeleisen, J., 1958. The significance of the product and sum rules to isotope fractionation processes. *Journal Name: Proc. of the Intern. Symposium on Isotope Separation; Other Information: Orig. Receipt Date: 31-DEC-58*, Medium: X; Size: Pages: 121-57.

Bigeleisen, J. and Goeppert-Mayer, M. G., 1947. Calculation of equilibrium constants for isotopic exchange reactions. *The Journal of Chemical Physics* **15**, 261-267.

Bigeleisen, J. and Wolfsberg, M., 1958. Theoretical and Experimental Aspects of Isotope Effects in Chemical Kinetics, *Advances in Chemical Physics*. John Wiley & Sons, Inc.

Blanchard, M., Poitrasson, F., Méheut, M., Lazzeri, M., Mauri, F., and Balan, E., 2009. Iron isotope fractionation between pyrite (FeS2), hematite (Fe2O3) and siderite (FeCO3): A first-principles density functional theory study. *Geochimica et Cosmochimica Acta* **73**, 6565-6578.

Brantley, S. L., Liermann, L. J., Guynn, R. L., Anbar, A., Icopini, G. A., and Barling, J., 2004. Fe isotopic fractionation during mineral dissolution with and without bacteria. *Geochimica et Cosmochimica Acta* **68**, 3189-3204.

Busigny, V. and Dauphas, N., 2007. Tracing paleofluid circulations using iron isotopes: A study of hematite and goethite concretions from the Navajo Sandstone (Utah, USA). *Earth and Planetary Science Letters* **254**, 272-287.

Chan, M. A., Beitler, B., Parry, W. T., Ormo, J., and Komatsu, G., 2004. A possible terrestrial analogue for haematite concretions on Mars. *Nature* **429**, 731-734.

Chan, M. A., Johnson, C. M., Beard, B. L., Bowman, J. R., and Parry, W. T., 2006. Iron isotopes constrain the pathways and formation mechanisms of terrestrial oxide concretions: A tool for tracing iron cycling on Mars? *Geosphere* **2**, 324-332.

Chumakov, A. I. and Sturhahn, W., 1999. Experimental aspects of inelastic nuclear resonance scattering. *Hyperfine Interactions* **123-124**, 781-808.

Cornell, R. M. and Schwertmann, U., 2003. *The Iron Oxides: Structure, Properties, Reactions, Occurences and Uses*. Wiley, New-York.

Craddock, P. R. and Dauphas, N., 2011. Iron and carbon isotope evidence for microbial iron respiration throughout the Archean. *Earth and Planetary Science Letters* **303**, 121-132.

Crosby, H., Roden, E., Johnson, C., and Beard, B., 2007a. Mechanisms of Fe isotope fractionation during dissimilatory Fe(III) reduction by S-putrefaciens and G-sulfurreducens. *Geochimica et Cosmochimica Acta* **71**, A194-A194.

Crosby, H. A., Johnson, C. M., Roden, E. E., and Beard, B. L., 2005. Coupled Fe(II)-Fe(III) electron and atom exchange as a mechanism for Fe isotope fractionation during dissimilatory iron oxide reduction. *Environmental Science & Technology* **39**, 6698-6704.

Crosby, H. A., Roden, E. E., Johnson, C. M., and Beard, B. L., 2007b. The mechanisms of iron isotope fractionation produced during dissimilatory Fe(III) reduction by Shewanella putrefaciens and Geobacter sulfurreducens. *Geobiology* **5**, 169-189.

Dauphas, N., Cates, N. L., Mojzsis, S. J., and Busigny, V., 2007a. Identification of chemical sedimentary protoliths using iron isotopes in the > 3750 Ma Nuvvuagittuq supracrustal belt, Canada. *Earth and Planetary Science Letters* **254**, 358-376.

Dauphas, N. and Morris, R. V., 2008. Iron isotopes in products of acid-sulfate basalt alteration: A prospective study for mars. *Meteoritics & Planetary Science* **43**, A35-A35.

Dauphas, N. and Rouxel, O., 2006. Mass spectrometry and natural variations of iron isotopes. *Mass Spectrometry Reviews* **25**, 515-550.

Dauphas, N., van Zuilen, M., Busigny, V., Lepland, A., Wadhwa, M., and Janney, P. E., 2007b. Iron isotope, major and trace element characterization of early Archean supracrustal rocks from SW Greenland: Protolith identification and metamorphic overprint. *Geochimica et Cosmochimica Acta* **71**, 4745-4770.

Dauphas, N., van Zuilen, M., Wadhwa, M., Davis, A. M., Marty, B., and Janney, P. E., 2004. Clues from Fe isotope variations on the origin of early Archean BIFs from Greenland. *Science* **306**, 2077-2080.

De Grave E., Van Alboom A. (1991) Evaluation of ferrous and ferric Mossbauer fractions. Phys. Chem. Minerals 18, 337-342.

De Grave E., Persoons R.M., Vandenberghe R.E., De Bakker P.M.A. (1993) Mossbauer study of the high-temperature phase of Co-substituted magnetites $Co_xFe_{3-x}O_4$. I. x<0.04. Phys. Rev., B47, 5881-5893.

Domagal-Goldman, S. D. and Kubicki, J. D., 2008. Density functional theory predictions of equilibrium isotope fractionation of iron due to redox changes and organic complexation. *Geochimica et Cosmochimica Acta* **72**, 5201-5216.

Domagal-Goldman, S. D., Paul, K. W., Sparks, D. L., and Kubicki, J. D., 2009. Quantum chemical study of the Fe(III)-desferrioxamine B siderophore complex--Electronic structure, vibrational frequencies, and equilibrium Fe-isotope fractionation. *Geochimica et Cosmochimica Acta* **73**, 1-12.

Dubiel, S. M., Cieslak, J., Sturhahn, W., Sternik, M., Piekarz, P., Stankov, S., and Parlinski, K., 2010. Vibrational Properties of alpha - and sigma -Phase Fe-Cr Alloy. *Physical Review Letters* **104**, 155503.

Elcombe M.M. (1974) Determination of atomic eigenvector magnitudes by isotopic substitution. J. Phys. C: Solid State Phys. 7, L202-L205.

Elcombe M.M., Hulston J.R. (1975) Calculation of sulphur isotope fractionation between sphalerite and galena using lattice dynamics. Earth and Planetary Science Letters 28, 172-180.

Fernandez-Remolar, D. C., Morris, R. V., Gruener, J. E., Amils, R., and Knoll, A. H., 2005. The RÌo Tinto Basin, Spain: Mineralogy, sedimentary geobiology, and implications for interpretation of outcrop rocks at Meridiani Planum, Mars. *Earth and Planetary Science Letters* **240**, 149-167.

Frost, C., von Blanckenburg, F., Schoenberg, R., Frost, B., and Swapp, S., 2007. Preservation of Fe isotope heterogeneities during diagenesis and metamorphism of banded iron formation. *Contributions to Mineralogy and Petrology* **153**, 211-235.

Golden, D. C., Ming, D. W., Morris, R. V., and Graff, T. G., 2008. Hydrothermal synthesis of hematite spherules and jarosite: Implications for diagenesis and hematite spherule formation in sulfate outcrops at Meridiani Planum, Mars. *American Mineralogist* **93**, 1201-1214.

Heimann, A., Johnson, C. M., Beard, B. L., Valley, J. W., Roden, E. E., Spicuzza, M. J., and Beukes, N. J., 2010. Fe, C, and O isotope compositions of banded iron formation carbonates demonstrate a major role for dissimilatory iron reduction in ~2.5†Ga marine environments. *Earth and Planetary Science Letters* **294**, 8-18.

Herzfeld K.F., Teller E., 1938. The vapor pressure of isotopes. Physical Review 54, 912-915.

Hill, P. S. and Schauble, E. A., 2008. Modeling the effects of bond environment on equilibrium iron isotope fractionation in ferric aquo-chloro complexes. *Geochimica et Cosmochimica Acta* **72**, 1939-1958.

Hill, P. S., Schauble, E. A., Shahar, A., Tonui, E., and Young, E. D., 2009. Experimental studies of equilibrium iron isotope fractionation in ferric aquo-chloro complexes. *Geochimica et Cosmochimica Acta* **73**, 2366-2381.

Hill, P. S., Schauble, E. A., and Young, E. D., 2010. Effects of changing solution chemistry on $Fe^{3+}/Fe^{2+}$ isotope fractionation in aqueous Fe-Cl solutions. *Geochimica et Cosmochimica Acta* **74**, 6669-6689.

Hu, M., Sturhahn, W., Toellner, T.S., Mannheim, P.D., Brown, D.E.,Zhao, J., Alp, E. (2003) Measuring velocity of sound with nuclear inelastic x-ray scattering. Physical Review B 67, 094304.

Icopini, G. A., Anbar, A. D., Ruebush, S. S., Tien, M., and Brantley, S. L., 2004. Iron isotope fractionation during microbial reduction of iron: The importance of adsorption. *Geology* **32**, 205-208.

Jackson, J. M., Hamecher, E. A., and Sturhahn, W., 2009. Nuclear resonant X-ray spectroscopy of $(Mg,Fe)SiO_3$ orthoenstatites. *Eur J Mineral* **21**, 551-560.

Jang, J.-H., Mathur, R., Liermann, L. J., Ruebush, S., and Brantley, S. L., 2008. An iron isotope signature related to electron transfer between aqueous ferrous iron and goethite. *Chemical Geology* **250**, 40-48.

Johnson, C., Beard, B., Beukes, N., Klein, C., and O'Leary, J., 2003. Ancient geochemical cycling in the Earth as inferred from Fe isotope studies of banded iron formations from the Transvaal Craton. *Contributions to Mineralogy and Petrology* **144**, 523-547.

Johnson, C. M., Beard, B. L., and Roden, E. E., 2008. The iron isotope fingerprints of redox and biogeochemical cycling in the modern and ancient Earth. *Annual Review of Earth and Planetary Sciences* **36**, 457-493.

Klingelhöfer, G., Morris, R. V., Bernhardt, B., Schröder, C., Rodionov, D. S., de Souza, P. A., Yen, A., Gellert, R., Evlanov, E. N., Zubkov, B., Foh, J., Bonnes, U., Kankeleit, E., Gütlich, P., Ming, D. W., Renz, F., Wdowiak, T., Squyres, S. W., and Arvidson, R. E., 2004. Jarosite and Hematite at Meridiani Planum from Opportunity's M√∂ssbauer Spectrometer. *Science* **306**, 1740-1745.

Kobayashi, H., Kamimura, T., Alf, egrave, Dario, Sturhahn, W., Zhao, J., and Alp, E. E., 2004. Phonon Density of States and Compression Behavior in Iron Sulfide under Pressure. *Physical Review Letters* **93**, 195503.

Kobayashi, H., Umemura, J., Kazekami, Y., Sakai, N., Alfé, D., Ohishi, Y., and Yoda, Y., 2007. Pressure-induced amorphization of CuFe S2 studied by Fe57 nuclear resonant inelastic scattering. *Physical Review B* **76**, 134108.

Kohn, V. G. and Chumakov, A. I., 2000. DOS: Evaluation of phonon density of states from nuclear resonant inelastic absorption. *Hyperfine Interactions* **125**, 205-221.

Lübbers, R., Grünsteudel, H. F., Chumakov, A. I., and Wortmann, G., 2000. Density of Phonon States in Iron at High Pressure. *Science* **287**, 1250-1253.

Lin, J.-F., Fei, Y., Sturhahn, W., Zhao, J., Mao, H.-k., and Hemley, R. J., 2004. Magnetic transition and sound velocities of Fe3S at high pressure: implications for Earth and planetary cores. *Earth and Planetary Science Letters* **226**, 33-40.

Lin, J.-F., Jacobsen, S. D., Sturhahn, W., Jackson, J. M., Zhao, J., and Yoo, C.-S., 2006. Sound velocities of ferropericlase in the Earth's lower mantle. *Geophys. Res. Lett.* **33**, L22304.

Lin, J.-F., Mao, Z., Yavas, H., Zhao, J., and Dubrovinsky, L., 2010. Shear wave anisotropy of textured hcp-Fe in the Earth's inner core. *Earth and Planetary Science Letters* **298**, 361-366.

Lin, J.-F., Sturhahn, W., Zhao, J., Shen, G., Mao, H.-k., and Hemley, R. J., 2005. Sound Velocities of Hot Dense Iron: Birch's Law Revisited. *Science* **308**, 1892-1894.

Lipkin, H. J., 1995. Mössbauer sum rules for use with synchrotron sources. *Physical Review B* **52**, 10073.

Lipkin, H. J., 1999. Mössbauer sum rules for use with synchrotron sources. *Hyperfine Interactions* **123-124**, 349-366.

Matsuhisa, Y., Goldsmith, J. R., and Clayton, R. N., 1978. Mechanisms of hydrothermal crystallization of quartz at 250¬∞C and 15 kbar. *Geochimica et Cosmochimica Acta* **42**, 173-182.

McLennan, S. M., Bell, J. F., Calvin, W. M., Christensen, P. R., Clark, B. C., de Souza, P. A., Farmer, J., Farrand, W. H., Fike, D. A., Gellert, R., Ghosh, A., Glotch, T. D., Grotzinger, J. P., Hahn, B., Herkenhoff, K. E., Hurowitz, J. A., Johnson, J. R., Johnson, S. S., Jolliff, B., Klingelhofer, G., Knoll, A. H., Learner, Z., Malin, M. C., McSween, H. Y., Pocock, J., Ruff, S. W., Soderblom, L. A., Squyres, S. W., Tosca, N. J., Watters, W. A., Wyatt, M. B., and Yen, A., 2005. Provenance and diagenesis of the evaporite-bearing Burns formation, Meridiani Planum, Mars. *Earth and Planetary Science Letters* **240**, 95-121.

Menéndez J., Page J.B., Guha S. (1994) The isotope effect on the Raman spectrum of molecular C60. Philosophical Magazine B 70, 651-659.

Mikutta, C., Wiederhold, J. G., Cirpka, O. A., Hofstetter, T. B., Bourdon, B., and Gunten, U. V., 2009. Iron isotope fractionation and atom exchange during sorption of ferrous iron to mineral surfaces. *Geochimica et Cosmochimica Acta* **73**, 1795-1812.

Morris, R. V., Ming, D. W., Graff, T. G., Arvidson, R. E., Bell Iii, J. F., Squyres, S. W., Mertzman, S. A., Gruener, J. E., Golden, D. C., Le, L., and Robinson, G. A., 2005. Hematite spherules in basaltic tephra altered under aqueous, acid-sulfate conditions on Mauna Kea volcano, Hawaii: Possible clues for the occurrence of hematite-rich spherules in the Burns formation at Meridiani Planum, Mars. *Earth and Planetary Science Letters* **240**, 168-178.

Morris, R. V., G. Klingelhöfer, C. Schröder, D. S. Rodionov, A. Yen, D. W. Ming, P. A. de Souza Jr., I. Fleischer, T. Wdowiak, R. Gellert, B. Bernhardt, E. N. Evlanov, B. Zubkov, J. Foh, U. Bonnes, E. Kankeleit, P. Gütlich, F. Renz, S. W. Squyres, and R. E. Arvidson (2006), Mössbauer mineralogy of rock, soil, and dust at Gusev Crater, Mars: Spirit's journey through weakly altered olivine basalt on the Plains and pervasively altered basalt in the Columbia Hills, J. Geophys. Res., 111, E02S13, doi:10.1029/2005JE002584.

Morris, R. V., G. Klingelhöfer, C. Schröder, I. Fleischer, D. W. Ming, A. S. Yen, R. Gellert, R. E. Arvidson, D. S. Rodionov, L. S. Crumpler, B. C. Clark, B. A. Cohen, T. J. McCoy, D. W. Mittlefehldt, M. E. Schmidt, P. A. de Souza Jr., and S. W. Squyres (2008), Iron mineralogy and aqueous alteration from Husband Hill through Home Plate at Gusev Crater, Mars: Results from the Mössbauer instrument on the Spirit Mars Exploration Rover, J. Geophys. Res., 113, E12S42, doi:10.1029/2008JE003201.

Nordstrom Darrell, K., Jenne Everett, A., and Ball James, W., 1979. Redox equilibria of iron in acid mine waters, in *Chemical Modeling in Aqueous Systems. ACS Symposium Series* **9**, 51-79.

Ottonello, G. and Vetuschi Zuccolini, M., 2008. The iron-isotope fractionation dictated by the carboxylic functional: An ab-initio investigation. *Geochimica et Cosmochimica Acta* **72**, 5920-5934.

Ottonello, G. and Zuccolini, M. V., 2009. Ab-initio structure, energy and stable Fe isotope equilibrium fractionation of some geochemically relevant H-O-Fe complexes. *Geochimica et Cosmochimica Acta* **73**, 6447-6469.

Peek, D. A., Fujita, I., Schmidt, M. C., and Simmons, R. O., 1992. Single-particle kinetic energies in solid neon. *Physical Review B* **45**, 9680-9687.

Poitrasson, F., Halliday, A. N., Lee, D. C., Levasseur, S., and Teutsch, N., 2004. Iron isotope differences between Earth, Moon, Mars and Vesta as possible records of contrasted accretion mechanisms. *Earth and Planetary Science Letters* **223**, 253-266.

Poitrasson, F., Roskosz, M., and Corgne, A., 2009. No iron isotope fractionation between molten alloys and silicate melt to 2000 °C and 7.7 GPa: Experimental evidence and implications for planetary differentiation and accretion. *Earth and Planetary Science Letters* **278**, 376-385.

Polyakov, V. B., 1997. Equilibrium fractionation of the iron isotopes: Estimation from Mössbauer spectroscopy data. *Geochimica et Cosmochimica Acta* **61**, 4213-4217.

Polyakov, V. B., 2009. Equilibrium Iron Isotope Fractionation at Core-Mantle Boundary Conditions. *Science* **323**, 912-914.

Polyakov, V. B., Clayton, R. N., Horita, J., and Mineev, S. D., 2007. Equilibrium iron isotope fractionation factors of minerals: Reevaluation from the data of nuclear inelastic resonant X-ray scattering and Mössbauer spectroscopy. *Geochimica et Cosmochimica Acta* **71**, 3833-3846.

Polyakov, V. B. and Mineev, S. D., 2000. The use of Mössbauer spectroscopy in stable isotope geochemistry. *Geochimica et Cosmochimica Acta* **64**, 849-865.

Polyakov, V. B., Mineev, S. D., Clayton, R. N., Hu, G., and Mineev, K. S., 2005a. Determination of tin equilibrium isotope fractionation factors from synchrotron radiation experiments. *Geochimica et Cosmochimica Acta* **69**, 5531-5536.

Polyakov, V.B., Mineev, S.D., Clayton, R.N., Hu, G., Gurevich, V.M., Khramov, D.A., Gavrichev, K.S., Gorbunov, V.E., Golushina, L.N., 2005b. Oxygen isotope fractionation factors involving cassiterite ($SnO_2$): I. Calculation of reduced partition function ratios from heat capacity and X-ray resonant studies. Geochimica et Cosmochimica Acta 69, 1287-1300.

Polyakov, V. B. and Soultanov, D. M., 2011. New data on equilibrium iron isotope fractionation among sulfides: Constraints on mechanisms of sulfide formation in hydrothermal and igneous systems. *Geochimica et Cosmochimica Acta* **75**, 1957-1974.

Röhlsberger R., Gersau E., Rüffer R., Sturhahn W., Toellner T.S., Chumakov A.I., Alp E.E. (1997) X-ray optics for μeV-resolved spectroscopy. Nuclear Instruments and Methods in Physics Research Section A 394, 251-255.

Rouxel, O., Dobbek, N., Ludden, J., and Fouquet, Y., 2003. Iron isotope fractionation during oceanic crust alteration. *Chemical Geology* **202**, 155-182.

Rustad, J. R., Casey, W. H., Yin, Q.-Z., Bylaska, E. J., Felmy, A. R., Bogatko, S. A., Jackson, V. E., and Dixon, D. A., 2010. Isotopic fractionation of $Mg^{2+}(aq)$, $Ca^{2+}(aq)$, and $Fe^{2+}(aq)$ with carbonate minerals. *Geochimica et Cosmochimica Acta* **74**, 6301-6323.

Rustad, J. R. and Dixon, D. A., 2009. Prediction of Iron-Isotope Fractionation Between Hematite (Œ±-Fe2O3) and Ferric and Ferrous Iron in Aqueous Solution from Density Functional Theory. *The Journal of Physical Chemistry A* **113**, 12249-12255.

Rustad, J. R. and Yin, Q.-Z., 2009. Iron isotope fractionation in the Earth/'s lower mantle. *Nature Geosci* **2**, 514-518.

Saunier, G. l., Pokrovski, G. S., and Poitrasson, F., 2011. First experimental determination of iron isotope fractionation between hematite and aqueous solution at hydrothermal conditions. *Geochimica et Cosmochimica Acta* **75**, 6629-6654.

Schauble, E. A., Rossman, G. R., and Taylor, H. P., 2001. Theoretical estimates of equilibrium Fe-isotope fractionations from vibrational spectroscopy. *Geochimica et Cosmochimica Acta* **65**, 2487-2497.

Schuessler, J. A., Schoenberg, R., Behrens, H., and Blanckenburg, F. v., 2007. The experimental calibration of the iron isotope fractionation factor between pyrrhotite and peralkaline rhyolitic melt. *Geochimica et Cosmochimica Acta* **71**, 417-433.

Seto, M., Yoda, Y., Kikuta, S., Zhang, X.W., Ando, M. (1995) Observation of nuclear resonant scattering accompanied by phonon excitation using synchrotron radiation. Phys. Rev. Lett. 74, 3828-3831.

Seto, M., Kitao, S., Kobayashi, Y., Haruki, R., Yoda, Y., Mitsui, T., and Ishikawa, T., 2003. Site-Specific Phonon Density of States Discerned using Electronic States. *Physical Review Letters* **91**, 185505.

Shahar, A., Young, E. D., and Manning, C. E., 2008. Equilibrium high-temperature Fe isotope fractionation between fayalite and magnetite: An experimental calibration. *Earth and Planetary Science Letters* **268**, 330-338.

Shen, G., Sturhahn, W., Alph, E. E., Zhao, J., Tollenner, T. S., Prakapenka, V. B., Meng, Y., and Mao, H. R., 2004. Phonon density of states in iron at high pressures and high temperatures. *Physics and Chemistry of Minerals* **31**, 353-359.

Skulan, J. L., Beard, B. L., and Johnson, C. M., 2002. Kinetic and equilibrium Fe isotope fractionation between aqueous Fe(III) and hematite. *Geochimica et Cosmochimica Acta* **66**, 2995-3015.

Stumm, W. and Morgan, J. J., 1996. *Aquatic Chemistry: Chemical Equilibria and Rates in Natural Waters*. Wiley, New-York.

Sturhahn, W., 2000. CONUSS and PHOENIX: Evaluation of nuclear resonant scattering data. *Hyperfine Interactions* **125**, 149-172.

Sturhahn, W., 2004. Nuclear resonant spectroscopy. *Journal of Physics: Condensed Matter* **16**, S497.

Sturhahn, W., Alp, E. E., Quast, K. W., and Toellner, T., 1999. Lamb-Mossbauer factor and second-order Dopller shift of hematite, *Advanced Photon Source User Activity Report 1999*.

Sturhahn W., Kohn V.G. 1999. Theoretical aspects of incoherent nuclear resonant scattering. Hyperfine Interactions 123/124, 367-399.

Sturhahn, W. and Jackson, J. M., 2007. Geophysical application of nuclear resonant spectroscopy. *Geological Society of America Special Paper* **421**, 157-174.

Sturhahn, W., Toellner, T. S., Alp, E. E., Zhang, X., Ando, M., Yoda, Y., Kikuta, S., Seto, M., Kimball, C. W., and Dabrowski, B., 1995. Phonon Density of States Measured by Inelastic Nuclear Resonant Scattering. *Physical Review Letters* **74**, 3832.

Teutsch, N., von Gunten, U., Porcelli, D., Cirpka, O. A., and Halliday, A. N., 2005. Adsorption as a cause for iron isotope fractionation in reduced groundwater. *Geochimica et Cosmochimica Acta* **69**, 4175-4185.

Toellner, T. S., 2000. Monochromatization of synchrotron radiation for nuclear resonant scattering experiments. *Hyperfine Interactions* **125**, 3-28.

Tosca, N. J., McLennan, S. M., Lindsley, D. H., and Schoonen, M. A. A., 2005. Acid-sulfate weathering of synthetic Martian basalt: The acid fog model revisited. *J. Geophys. Res.* **109**, E05003.

Tsunoda, Y., Kurimoto, Y., Seto, M., Kitao, S., and Yoda, Y., 2002. Phonon density of states of gamma -Fe precipitates in Cu. *Physical Review B* **66**, 214304.

Wang, K., Moynier, F., Dauphas, N. Barrat, J.-A. Craddock, P., Sio, C.K. 2012. Iron isotope fractionation in planetary crusts. *Geochimica et Cosmochimica Acta*, in press.

Welch, S. A., Beard, B. L., Johnson, C. M., and Braterman, P. S., 2003. Kinetic and equilibrium Fe isotope fractionation between aqueous Fe(II) and Fe(III). *Geochimica et Cosmochimica Acta* **67**, 4231-4250.

Weyer, S., Anbar, A. D., Brey, G. P., Munker, C., Mezger, K., and Woodland, A. B., 2005. Iron isotope fractionation during planetary differentiation. *Earth and Planetary Science Letters* **240**, 251-264.

Wolfsberg, M., 1969. Isotope effects. *Annual Review of Physical Chemistry* **20**, 449-478.

Wu, L., Beard, B. L., Roden, E. E., and Johnson, C. M., 2011. Stable Iron Isotope Fractionation Between Aqueous Fe(II) and Hydrous Ferric Oxide. *Environmental Science & Technology* **45**, 1847-1852.

Young, E. D., Galy, A., and Nagahara, H., 2002. Kinetic and equilibrium mass-dependent isotope fractionation laws in nature and their geochemical and cosmochemical significance. *Geochimica et Cosmochimica Acta* **66**, 1095-1104.

1121 Zolotov, M. Y. and Shock, E. L., 2005. Formation of jarosite-bearing deposits through
1122     aqueous oxidation of pyrite at Meridiani Planum, Mars. *Geophys. Res. Lett.* **32**,
1123     L21203.
1124
1125
1126
1127

**Table(s)**

Table 1. $^{56}$Fe/$^{54}$Fe β-factors calculated from force constants, $1{,}000 \times \ln\beta = 2{,}904 \langle F\rangle/T^2 - B_2 \langle F\rangle^2/T^4$ with $\langle F\rangle$ in N/m and T in K. $B_2$ values are estimates based on g(E) or S(E) data available (actual or digitized values).
In some previous studies, the energy scan during NRIXS measurements may have been too narrow and some force constants may need to be reevaluated when new data becomes available.

| Phase | P, T, size | Force constant (N/m) | $B_2$ | $B_2$ from g or S, actual or digitized | $10^3 \times \ln\beta$ 22 °C | $10^3 \times \ln\beta$ 22 °C, $B_2$=52,000 | $10^3 \times \ln\beta$ 500 °C | References | Litterature data, $10^3$: 22 °C | 500 °C | References |
|---|---|---|---|---|---|---|---|---|---|---|---|
| Goethite | | 314.1 ± 13.9 | 61951 | S+a | 9.67 ± 0.39 | 9.8 | 1.51 ± 0.07 | This study | | | |
| Potassium-jarosite | | 264.5 ± 11.6 | 58394 | S+a | 8.28 ± 0.34 | 8.3 | 1.27 ± 0.06 | This study | | | |
| Hydronium-jarosite | | 309.7 ± 13.6 | 64918 | S+a | 9.50 ± 0.38 | 9.7 | 1.49 ± 0.06 | This study | | | |
| Deoxymyoglobin | | 174 ± 16 | 42068 | g+d | 5.63 ± 0.50 | 5.6 | 0.84 ± 0.08 | Adams et al. (2006) | | | |
| Metmyoglobin | | 245 ± 9 | 39704 | g+d | 7.85 ± 0.28 | 7.8 | 1.18 ± 0.04 | Adams et al. (2006) | | | |
| Cytochrome f, oxidized | | 313 ± 34 | 39280 | g+d | 9.93 ± 1.02 | 9.8 | 1.51 ± 0.16 | Adams et al. (2006) | | | |
| Cytochrome f, reduced | | 342 ± 18 | 36353 | g+d | 10.84 ± 0.54 | 10.6 | 1.65 ± 0.09 | Adams et al. (2006) | | | |
| Orthoenstatite (Mg$_{0.93}$,Fe$_{0.07}$)SiO$_3$ | | 195 ± 5 | 66123 | g+d | 6.17 ± 0.15 | 6.2 | 0.94 ± 0.02 | Jackson et al. (2009) | | | |
| Orthoenstatite (Mg$_{0.87}$,Fe$_{0.13}$)SiO$_3$ | | 170 ± 3 | 65493 | g+d | 5.42 ± 0.09 | 5.5 | 0.82 ± 0.01 | Jackson et al. (2009) | | | |
| Orthoenstatite (Mg$_{0.80}$,Fe$_{0.20}$)SiO$_3$ | | 165 ± 5 | 65442 | g+d | 5.27 ± 0.15 | 5.3 | 0.80 ± 0.02 | Jackson et al. (2009) | | | |
| Hematite | 5 K | 252 ± 4 | 54915 | g+a | 7.94 ± 0.12 | 8.0 | 1.21 ± 0.02 | Sturhahn et al. (1999) | | | |
| Hematite | 50 K | 248 ± 7 | 54915 | g+a | 7.82 ± 0.21 | 7.8 | 1.20 ± 0.03 | Sturhahn et al. (1999) | | | |
| Hematite | 100 K | 241 ± 6 | 54915 | g+a | 7.61 ± 0.18 | 7.6 | 1.16 ± 0.03 | Sturhahn et al. (1999) | | | |
| Hematite | 150 K | 249 ± 7 | 54915 | g+a | 7.85 ± 0.21 | 7.9 | 1.20 ± 0.03 | Sturhahn et al. (1999) | | | |
| Hematite | 200 K | 244 ± 8 | 54915 | g+a | 7.70 ± 0.24 | 7.7 | 1.18 ± 0.04 | Sturhahn et al. (1999) | | | |
| Hematite | 240 K | 245 ± 9 | 54915 | g+a | 7.73 ± 0.27 | 7.8 | 1.18 ± 0.04 | Sturhahn et al. (1999) | | | |
| Hematite | 260 K | 235 ± 9 | 54915 | g+a | 7.43 ± 0.27 | 7.5 | 1.13 ± 0.04 | Sturhahn et al. (1999) | | | |
| Hematite | 300 K | 239 ± 9 | 54915 | g+a | 7.55 ± 0.27 | 7.6 | 1.15 ± 0.04 | Sturhahn et al. (1999) | | | |
| Magnetite | | 230.48 ± 5.5 | 57163 | g+a | 7.28 0.16 | 7.3 | 1.11 0.03 | Alp, pers. comm. | | | |
| FeS, Troilite | 1.5 Gpa | 102 | 61060 | g+d | 3.32 | 3.3 | 0.49 | Kobayashi et al. (2004) | 3.33 | 0.50 | Polyakov & Soultanov (2011) |
| FeS, Troilite | 2.5 GPa | 102 | 61060 | g+d | 3.32 | 3.3 | 0.49 | Kobayashi et al. (2004) | | | |
| FeS, Troilite | 3 GPa | 109 ± 8 | 61060 | g+d | 3.54 ± 0.25 | 3.6 | 0.53 ± 0.04 | Kobayashi et al. (2004) | | | |
| FeS, MnP-type | 4 GPa | 98 ± 10 | 55909 | g+d | 3.20 ± 0.32 | 3.2 | 0.47 ± 0.05 | Kobayashi et al. (2004) | | | |
| FeS, MnP-type | 5 GPa | 97 ± 9 | 55909 | g+d | 3.16 ± 0.29 | 3.2 | 0.47 ± 0.04 | Kobayashi et al. (2004) | | | |
| FeS, MnP-type | 6 GPa | 120 ± 14 | 55909 | g+d | 3.89 ± 0.44 | 3.9 | 0.58 ± 0.07 | Kobayashi et al. (2004) | | | |
| FeS, monoclinic | 9.5 GPa | 173 ± 13 | 44277 | g+d | 5.59 ± 0.41 | 5.6 | 0.84 ± 0.06 | Kobayashi et al. (2004) | | | |
| FeS, monoclinic | 12 GPa | 173 ± 21 | 44277 | g+d | 5.59 ± 0.66 | 5.6 | 0.84 ± 0.10 | Kobayashi et al. (2004) | | | |
| CuFeS$_2$ Chalcopyrite | 0 Gpa | 146 ± 2 | 56616 | g+d | 4.71 ± 0.06 | 4.7 | 0.71 ± 0.01 | Kobayashi et al. (2007) | 5.4 | 0.8 | Polyakov & Soultanov (2011) |
| CuFeS$_2$ Chalcopyrite | 1 Gpa | 146 ± 3 | 56616 | g+d | 4.71 ± 0.09 | 4.7 | 0.71 ± 0.01 | Kobayashi et al. (2007) | | | |
| CuFeS$_2$ Chalcopyrite | 3 Gpa | 147 ± 2 | 56616 | g+d | 4.74 ± 0.06 | 4.8 | 0.71 ± 0.01 | Kobayashi et al. (2007) | | | |
| CuFeS$_2$ Chalcopyrite | 4.5 Gpa | 149 ± 2 | 58621 | g+d | 4.80 ± 0.06 | 4.8 | 0.72 ± 0.01 | Kobayashi et al. (2007) | | | |
| CuFeS$_2$ Chalcopyrite | 6 Gpa | 151 ± 2 | 58621 | g+d | 4.86 ± 0.06 | 4.9 | 0.73 ± 0.01 | Kobayashi et al. (2007) | | | |
| CuFeS$_2$ Amorphous | 8 Gpa | 141 ± 2 | 58621 | g+d | 4.55 ± 0.06 | 4.6 | 0.68 ± 0.01 | Kobayashi et al. (2007) | | | |
| CuFeS$_2$ Amorphous | 14 Gpa | 139 ± 3 | 56255 | g+d | 4.49 ± 0.09 | 4.5 | 0.67 ± 0.01 | Kobayashi et al. (2007) | | | |
| Fe$_3$S | 0 | 119.7 ± 6.6 | 53756 | g+d | 3.89 ± 0.21 | 3.9 | 0.58 ± 0.03 | Lin et al. (2004) | 3.6 | 0.5 | Polyakov & Soultanov (2011) |
| Fe$_3$S | 6.1 GPa | 175.9 ± 6 | 49944 | g+d | 5.66 ± 0.19 | 5.7 | 0.85 ± 0.03 | Lin et al. (2004) | | | |
| Fe$_3$S | 13.1 GPa | 198.9 ± 5 | 47057 | g+d | 6.39 ± 0.15 | 6.4 | 0.96 ± 0.02 | Lin et al. (2004) | | | |
| Fe$_3$S | 21.3 GPa | 213.1 ± 4.9 | 44568 | g+d | 6.84 ± 0.15 | 6.8 | 1.03 ± 0.02 | Lin et al. (2004) | | | |
| Fe$_3$S | 28 GPa | 237 ± 12 | 44344 | g+d | 7.57 ± 0.37 | 7.5 | 1.14 ± 0.06 | Lin et al. (2004) | | | |
| Fe$_3$S | 37.7 GPa | 258.2 ± 5 | 43560 | g+d | 8.22 ± 0.15 | 8.2 | 1.25 ± 0.02 | Lin et al. (2004) | | | |
| Fe$_3$S | 45.2 GPa | 291.3 ± 6.2 | 43034 | g+d | 9.23 ± 0.19 | 9.1 | 1.40 ± 0.03 | Lin et al. (2004) | | | |
| Fe$_3$S | 57.1 GPa | 323.2 ± 7.8 | 41812 | g+d | 10.20 ± 0.23 | 10.1 | 1.56 ± 0.04 | Lin et al. (2004) | | | |
| (Mg$_{0.75}$, Fe$_{0.25}$)O | 0 GPa | 180 | 53934 | g+d | 5.77 | 5.8 | 0.87 | Lin et al. (2006) | 5.3 | 0.8 | Polyakov (2009) |
| (Mg$_{0.75}$, Fe$_{0.25}$)O | 8 Gpa | 203 | 51097 | g+d | 6.49 | 6.5 | 0.98 | Lin et al. (2006) | 6.1 | 1.0 | Polyakov (2009) |
| (Mg$_{0.75}$, Fe$_{0.25}$)O | 23 Gpa | 211 | 51097 | g+d | 6.75 | 6.7 | 1.02 | Lin et al. (2006) | | | |
| (Mg$_{0.75}$, Fe$_{0.25}$)O | 23 Gpa | 234 | 51097 | g+d | 7.44 | 7.4 | 1.13 | Lin et al. (2006) | | | |
| (Mg$_{0.75}$, Fe$_{0.25}$)O | 31 Gpa | 245 | 52392 | g+d | 7.75 | 7.8 | 1.18 | Lin et al. (2006) | 6.1 | 1.0 | Polyakov (2009) |
| (Mg$_{0.75}$, Fe$_{0.25}$)O | 42 Gpa | 245 | 48188 | g+d | 7.78 | 7.8 | 1.18 | Lin et al. (2006) | 7.3 | 1.2 | Polyakov (2009) |
| (Mg$_{0.75}$, Fe$_{0.25}$)O | 52 Gpa | 306 | 50378 | g+d | 9.57 | 9.5 | 1.47 | Lin et al. (2006) | 8.8 | 1.4 | Polyakov (2009) |
| (Mg$_{0.75}$, Fe$_{0.25}$)O | 58 Gpa | 360 | 50378 | g+d | 11.14 | 11.1 | 1.73 | Lin et al. (2006) | | | |
| (Mg$_{0.75}$, Fe$_{0.25}$)O | 62 Gpa | 368 | 49932 | g+d | 11.39 | 11.4 | 1.77 | Lin et al. (2006) | 10.6 | 1.7 | Polyakov (2009) |
| (Mg$_{0.75}$, Fe$_{0.25}$)O | 74 Gpa | 400 | 49932 | g+d | 12.28 | 12.2 | 1.92 | Lin et al. (2006) | | | |
| (Mg$_{0.75}$, Fe$_{0.25}$)O | 92 Gpa | 477 | 41983 | g+d | 14.65 | 14.3 | 2.29 | Lin et al. (2006) | 12.8 | 2.0 | Polyakov (2009) |
| (Mg$_{0.75}$, Fe$_{0.25}$)O | 109 Gpa | 666 | 38115 | g+d | 19.96 | 19.2 | 3.19 | Lin et al. (2006) | 13.7 | 2.1 | Polyakov (2009) |
| α-Fe (bcc) | | 175.15 ± 2.1 | 37089 | g+a | 5.69 ± 0.07 | 5.6 | 0.85 ± 0.01 | Alp et al. (2001) | 5.6 | 0.8 | Polyakov et al. (2007) |
| γ-Fe (fcc) precipitate in Cu | 3 nm | 132 ± 9 | 37639 | g+d | 4.31 ± 0.29 | 4.3 | 0.64 ± 0.04 | Tsunoda et al. (2002) | | | |
| γ-Fe (fcc) precipitate in Cu | 8 nm | 134 ± 10 | 36015 | g+d | 4.38 ± 0.32 | 4.3 | 0.65 ± 0.05 | Tsunoda et al. (2002) | | | |
| γ-Fe (fcc) precipitate in Cu | 15 nm | 138 ± 8 | 35395 | g+d | 4.51 ± 0.26 | 4.5 | 0.67 ± 0.04 | Tsunoda et al. (2002) | | | |
| γ-Fe (fcc) precipitate in Cu | 30 nm | 139 ± 8 | 35705 | g+d | 4.54 ± 0.26 | 4.5 | 0.67 ± 0.04 | Tsunoda et al. (2002) | | | |
| γ-Fe (fcc) precipitate in Cu | 50 nm | 147 ± 9 | 36299 | g+d | 4.80 ± 0.29 | 4.8 | 0.71 ± 0.04 | Tsunoda et al. (2002) | | | |
| γ-Fe (fcc) precipitate in Cu | 80 nm | 139 ± 10 | 36138 | g+d | 4.54 ± 0.32 | 4.5 | 0.67 ± 0.05 | Tsunoda et al. (2002) | | | |
| ε-Fe (hcp) | 36 Gpa, 300 K | 291.2 ± 8.7 | 35773 | g+d | 9.31 ± 0.27 | 9.1 | 1.41 ± 0.04 | Lin et al. (2005) | | | |
| ε-Fe (hcp) | 43.3 Gpa, 300 K | 263.2 ± 5.2 | 35773 | g+d | 8.45 ± 0.16 | 8.3 | 1.27 ± 0.02 | Lin et al. (2005) | | | |
| ε-Fe (hcp) | 44 Gpa, 300 K | 276.6 ± 6 | 35773 | g+d | 8.86 ± 0.18 | 8.7 | 1.34 ± 0.03 | Lin et al. (2005) | | | |
| ε-Fe (hcp) | 50.5 Gpa, 300 K | 296 ± 7.6 | 35773 | g+d | 9.45 ± 0.23 | 9.3 | 1.43 ± 0.04 | Lin et al. (2005) | | | |
| ε-Fe (hcp) | 54.7 Gpa, 300 K | 308.6 ± 9.9 | 35773 | g+d | 9.84 ± 0.30 | 9.6 | 1.49 ± 0.05 | Lin et al. (2005) | | | |
| ε-Fe (hcp) | 51.1 Gpa, 300 K | 312.8 ± 8.7 | 35773 | g+d | 9.97 ± 0.26 | 9.8 | 1.51 ± 0.04 | Lin et al. (2005) | | | |
| ε-Fe (hcp) | 54.6 Gpa, 300 K | 325.6 ± 5.9 | 35773 | g+d | 10.35 ± 0.18 | 10.1 | 1.57 ± 0.03 | Lin et al. (2005) | | | |
| ε-Fe (hcp) | 60 Gpa, 300 K | 329.5 ± 9.9 | 35773 | g+d | 10.47 ± 0.30 | 10.2 | 1.59 ± 0.05 | Lin et al. (2005) | | | |
| ε-Fe (hcp) | 71 Gpa, 300 K | 328.6 ± 4.9 | 35773 | g+d | 10.45 ± 0.15 | 10.2 | 1.59 ± 0.02 | Lin et al. (2005) | | | |
| ε-Fe (hcp) | 72.8 Gpa, 700 K | 299.3 ± 14 | 35773 | g+d | 9.56 ± 0.43 | 9.4 | 1.45 ± 0.07 | Lin et al. (2005) | | | |
| ε-Fe (hcp) | 73.3 Gpa, 880 K | 299.5 ± 21.5 | 35773 | g+d | 9.56 ± 0.66 | 9.4 | 1.45 ± 0.10 | Lin et al. (2005) | | | |
| ε-Fe (hcp) | 39 Gpa, 1000 K | 267.3 ± 7.8 | 35773 | g+d | 8.57 ± 0.24 | 8.4 | 1.29 ± 0.04 | Lin et al. (2005) | | | |
| ε-Fe (hcp) | 57.5 Gpa, 1000 K | 292.8 ± 7.7 | 35773 | g+d | 9.36 ± 0.24 | 9.2 | 1.41 ± 0.04 | Lin et al. (2005) | | | |
| ε-Fe (hcp) | 46.5 Gpa, 1100 K | 257 ± 10 | 38686 | g+d | 8.23 ± 0.31 | 8.1 | 1.24 ± 0.05 | Lin et al. (2005) | | | |
| ε-Fe (hcp) | 47.2 Gpa, 1100 K | 277.8 ± 18.5 | 38686 | g+d | 8.87 ± 0.56 | 8.7 | 1.34 ± 0.09 | Lin et al. (2005) | | | |
| ε-Fe (hcp) | 57.9 Gpa, 1200 K | 294.9 ± 13.1 | 38686 | g+d | 9.39 ± 0.40 | 9.2 | 1.42 ± 0.06 | Lin et al. (2005) | | | |
| ε-Fe (hcp) | 54.6 Gpa, 1300 K | 266.8 ± 22.9 | 38686 | g+d | 8.53 ± 0.70 | 8.4 | 1.29 ± 0.11 | Lin et al. (2005) | | | |
| ε-Fe (hcp) | 54.1 Gpa, 1500 K | 166.5 ± 8.2 | 38686 | g+d | 5.41 ± 0.26 | 5.4 | 0.81 ± 0.04 | Lin et al. (2005) | | | |
| ε-Fe (hcp) | 58.1 Gpa, 1700 K | 196.9 ± 15 | 38686 | g+d | 6.37 ± 0.47 | 6.3 | 0.95 ± 0.07 | Lin et al. (2005) | | | |
| ε-Fe (hcp) | 158 Gpa, c axis | 511.7 ± 5.6 | 33820 | g+d | 15.89 ± 0.16 | 15.3 | 2.46 ± 0.03 | Lin et al. (2010) | | | |
| ε-Fe (hcp) | 158 Gpa, 30 ° to a axis | 508.1 ± 3.5 | 33820 | g+d | 15.79 ± 0.10 | 15.2 | 2.44 ± 0.02 | Lin et al. (2010) | | | |
| ε-Fe (hcp) | 158 Gpa, a axis | 491.1 ± 4.5 | 33820 | g+d | 15.30 ± 0.13 | 14.7 | 2.36 ± 0.02 | Lin et al. (2010) | | | |
| ε-Fe (hcp) | 172 Gpa, c axis | 531.4 ± 8.8 | 33820 | g+d | 16.46 ± 0.25 | 15.8 | 2.55 ± 0.04 | Lin et al. (2010) | | | |
| ε-Fe (hcp) | 172 Gpa, 28 ° to a axis | 534.5 ± 7.7 | 33820 | g+d | 16.54 ± 0.22 | 15.9 | 2.57 ± 0.04 | Lin et al. (2010) | | | |
| ε-Fe (hcp) | 172 Gpa, a axis | 526.8 ± 8.8 | 33820 | g+d | 16.32 ± 0.25 | 15.7 | 2.53 ± 0.04 | Lin et al. (2010) | | | |
| α-Fe (hcp) | 0 Gpa | 185 ± 12 | 36246 | g+d | 6.00 ± 0.38 | 5.9 | 0.90 ± 0.06 | Lubbers et al. (2011) | | | |
| ε-Fe (hcp) | 20 Gpa | 320 ± 15 | 34950 | g+d | 10.20 ± 0.46 | 10.0 | 1.54 ± 0.07 | Lubbers et al. (2011) | | | |
| ε-Fe (hcp) | 32 Gpa | 365 ± 15 | 34950 | g+d | 11.55 ± 0.45 | 11.3 | 1.76 ± 0.07 | Lubbers et al. (2011) | | | |
| ε-Fe (hcp) | 42 Gpa | 388 ± 15 | 34950 | g+d | 12.24 ± 0.45 | 11.9 | 1.87 ± 0.07 | Lubbers et al. (2011) | | | |
| ε-Fe (hcp) | 29 Gpa, 300 K | 253 ± 17 | 38678 | g+d | 8.11 ± 0.52 | 8.0 | 1.22 ± 0.08 | Shen et al. (2004) | | | |
| ε-Fe (hcp) | 29 Gpa, 430 K | 250 ± 19 | 35552 | g+d | 8.04 ± 0.59 | 7.9 | 1.21 ± 0.09 | Shen et al. (2004) | | | |
| ε-Fe (hcp) | 29 Gpa, 720 K | 236 ± 15 | 33720 | g+d | 7.62 ± 0.47 | 7.5 | 1.14 ± 0.07 | Shen et al. (2004) | | | |
| α-Fe$_{52.5}$Cr$_{47.5}$ | 298 K | 156 ± 1 | 36329 | g+d | 5.08 ± 0.03 | 5.0 | 0.76 ± 0.00 | Dubiel et al. (2010) | | | |
| σ-Fe$_{52.5}$Cr$_{47.5}$ | 298 K | 157 ± 2 | 39792 | g+d | 5.10 ± 0.06 | 5.1 | 0.76 ± 0.01 | Dubiel et al. (2010) | | | |
| σ-Fe$_{52.5}$Cr$_{47.5}$ | 20 K | 155.1 ± 0.7 | 39792 | g+d | 5.04 ± 0.02 | 5.0 | 0.75 ± 0.00 | Dubiel et al. (2010) | | | |

**Table 2.** Thermoelastic properties derived from NRIXS data. Uncertainties in parentheses take into account systematic errors on the mean force constant from S(E).

| | | Goethite | H-Jarosite | K-Jarosite |
|---|---|---|---|---|
| *From S(E)* | Lamb-Moessbauer factor | 0.77 ± 0.00 | 0.65 ± 0.00 | 0.68 ± 0.00 |
| | Kinetic energy/atom (meV) | 45.43 ± 0.38 | 46.41 ± 0.38 | 45.98 ± 0.27 |
| | Mean force constant (N/m) | 306.87 ± 9.43 | 302.38 ± 9.02 | 262.04 ± 5.82 |
| *From S(E) after refinement* | Lamb-Moessbauer factor | 0.77 ± 0.00 | 0.65 ± 0.00 | 0.68 ± 0.00 |
| | Kinetic energy/atom (meV) | 45.54 ± 0.38 | 46.51 ± 0.38 | 46.00 ± 0.27 |
| | Mean force constant (N/m)* | 314.10 ± 9.66 (13.90) | 309.71 ± 9.24 (13.61) | 264.48 ± 5.87 (11.60) |
| *From PDOS* | Lamb-Moessbauer factor | 0.77 ± 0.00 | 0.66 ± 0.00 | 0.69 ± 0.00 |
| | Kinetic energy/atom (meV) | 45.32 ± 0.46 | 45.12 ± 0.52 | 44.50 ± 0.31 |
| | Mean force constant (N/m) | 316.57 ± 10.83 | 307.26 ± 12.67 | 270.83 ± 7.04 |
| | Lamb-Moessbauer factor at T=0 | 0.92 ± 0.00 | 0.91 ± 0.00 | 0.91 ± 0.00 |
| | Kinetic energy/atom at T=0 (meV) | 24.86 ± 0.40 | 23.78 ± 0.47 | 23.07 ± 0.28 |
| | Vibrational specific heat ($k_B$/atom) | 2.57 ± 0.01 | 2.59 ± 0.02 | 2.62 ± 0.01 |
| | Vibrational entropy ($k_B$/atom) | 2.92 ± 0.01 | 3.21 ± 0.01 | 3.18 ± 0.01 |
| | Critical temperature (K) | 1209.20 ± 5.80 | 745.00 ± 2.06 | 850.60 ± 1.87 |
| *Mathematica script* | Force constant from S(E) (N/m) | 307 ± 9 | 303 ± 9 | 263 ± 6 |
| | Force constant from g(E) (N/m) | 316 ± 11 | 307 ± 13 | 271 ± 7 |
| 1000 ln β coefficients from S(E)* | $A_1$ | 8.93E+05 ± 2.73E+04 | 8.80E+05 ± 2.61E+04 | 7.63E+05 ± 1.69E+04 |
| $A_1/T^2+A_2/T^4+A_3/T^6$ (Eq. 8) | $A_2$ | -7.35E+09 ± 6.93E+08 | -7.45E+09 ± 6.61E+08 | -6.64E+09 ± 3.95E+08 |
| with T in K | $A_3$ | 1.39E+14 ± 2.97E+13 | 1.38E+14 ± 2.86E+13 | 1.33E+14 ± 1.65E+13 |
| 1000 ln β coefficients from g(E) | $A_1$ | 9.19E+05 ± 3.16E+04 | 8.92E+05 ± 3.79E+04 | 7.86E+05 ± 2.11E+04 |
| $A_1/T^2+A_2/T^4+A_3/T^6$ (Eq. 7) | $A_2$ | -8.15E+09 ± 8.30E+08 | -8.06E+09 ± 1.01E+09 | -4.88E+09 ± 5.33E+08 |
| with T in K | $A_3$ | 1.96E+14 ± 3.59E+13 | 1.93E+14 ± 4.43E+13 | 6.45E+13 ± 2.29E+13 |
| $B_1<F>/T^2-B_2<F>^2/T^4$ (Eq. 17)* | $B_1$ | 2904.48 | 2904.48 | 2904.48 |
| with <F> in N/,m and T in K | $B_2$ | 61951.30 | 64917.80 | 58394.20 |

*Preferred methods.

Figure(s)

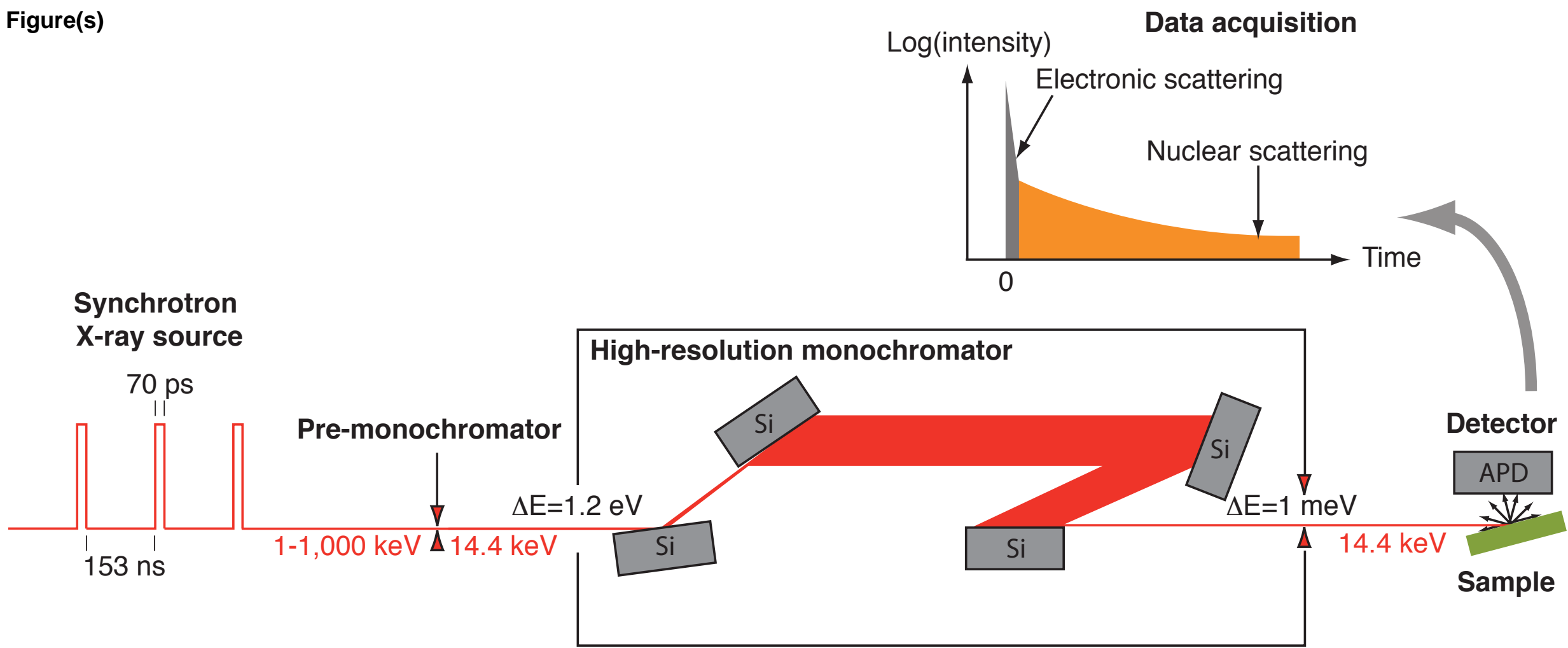

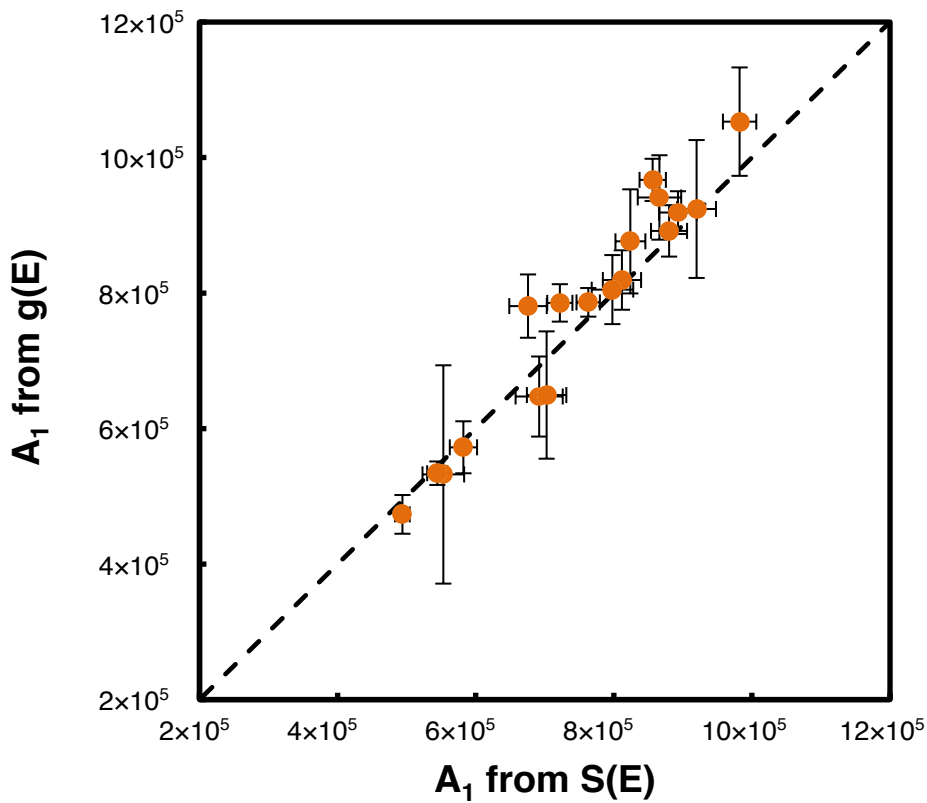
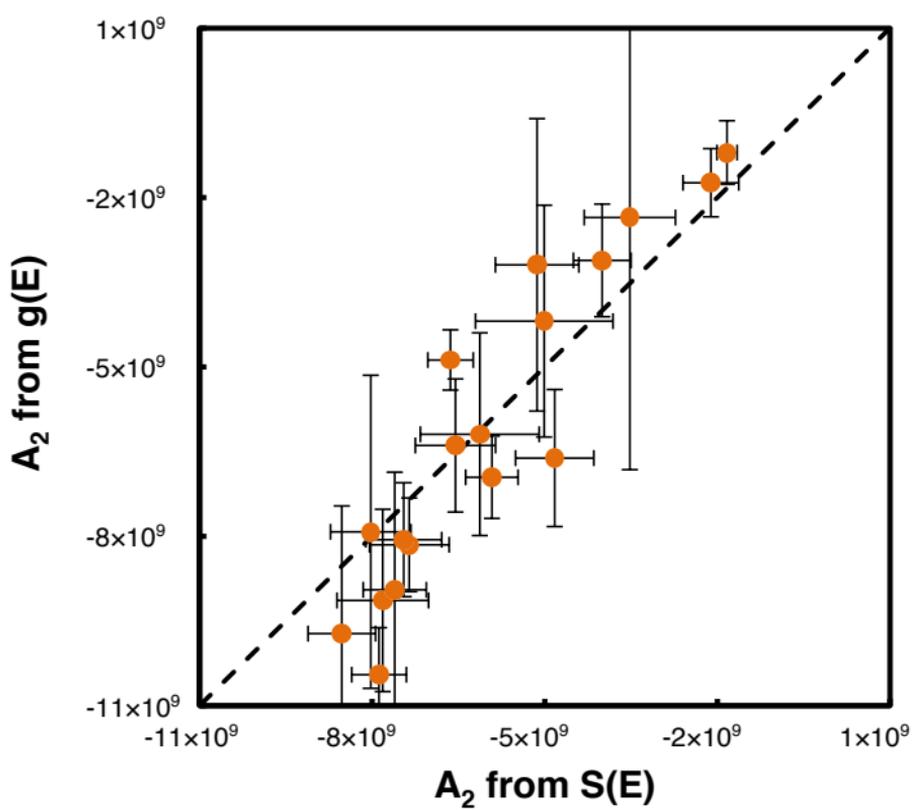
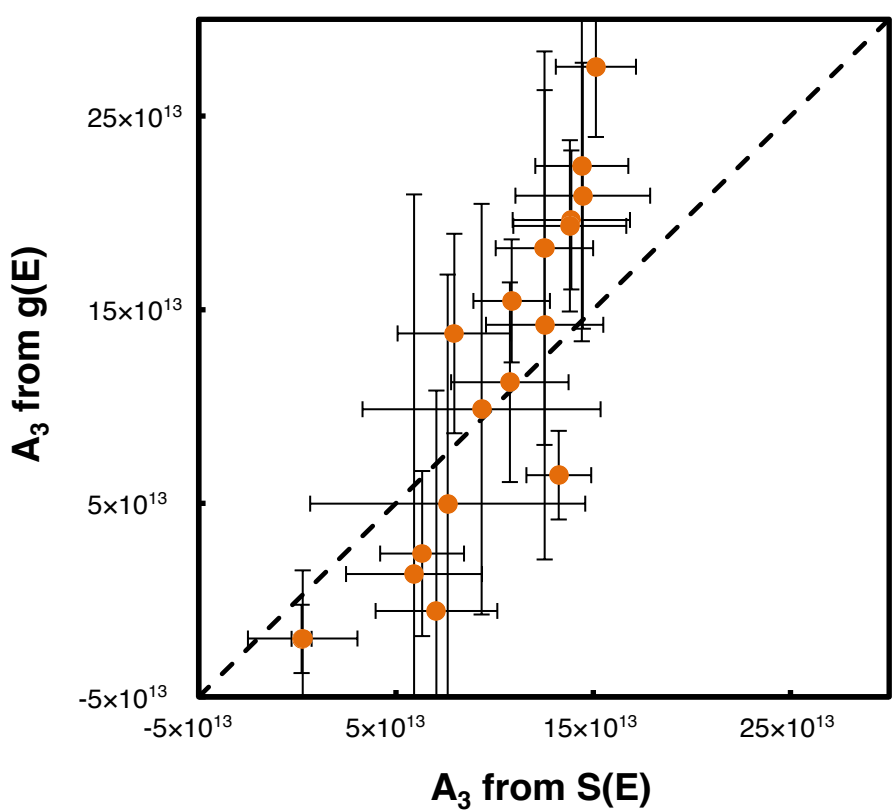

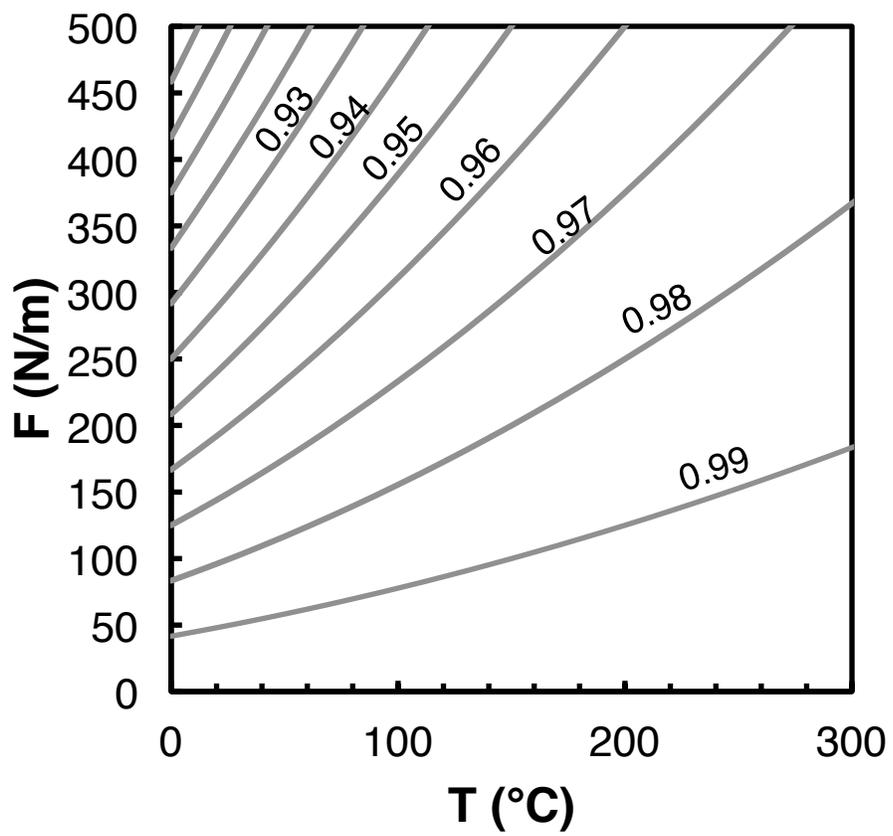

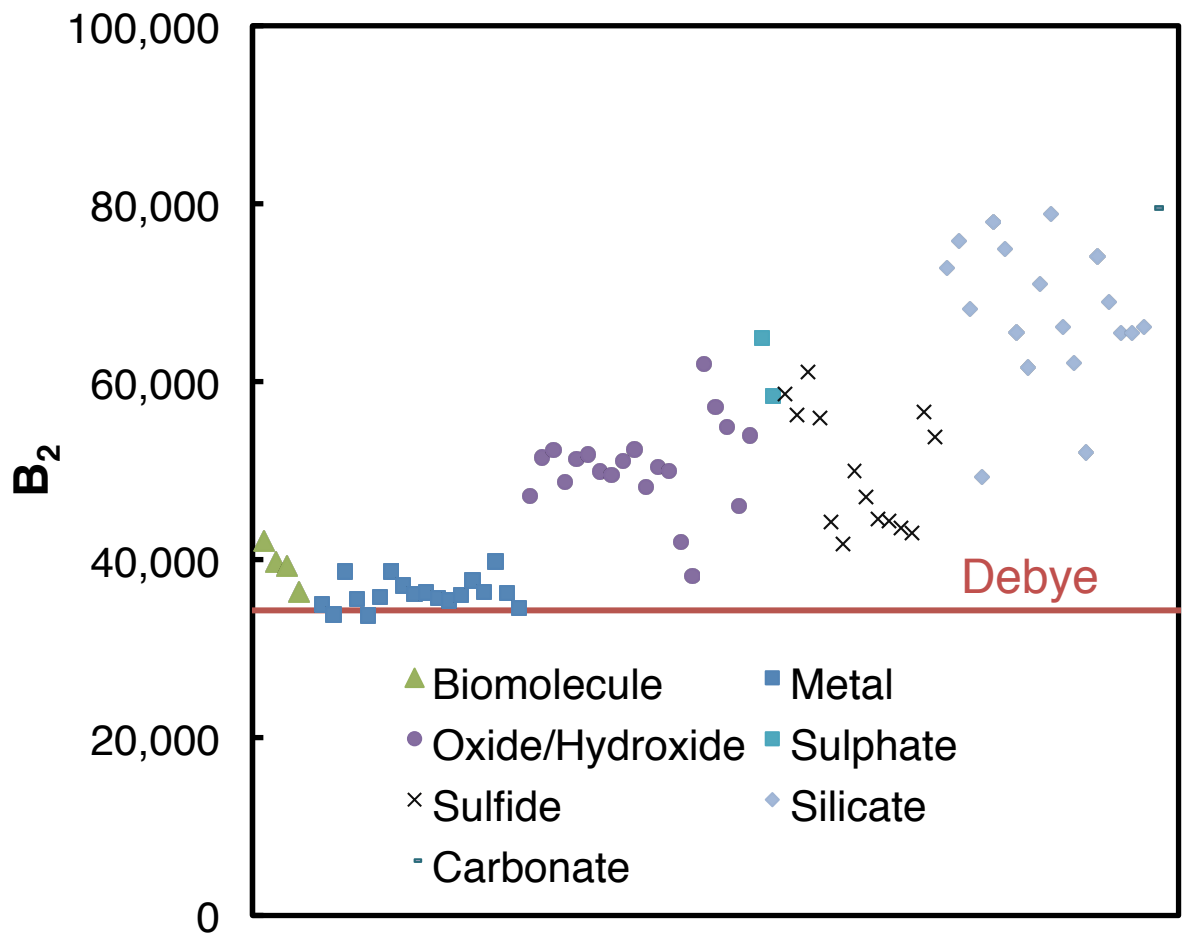

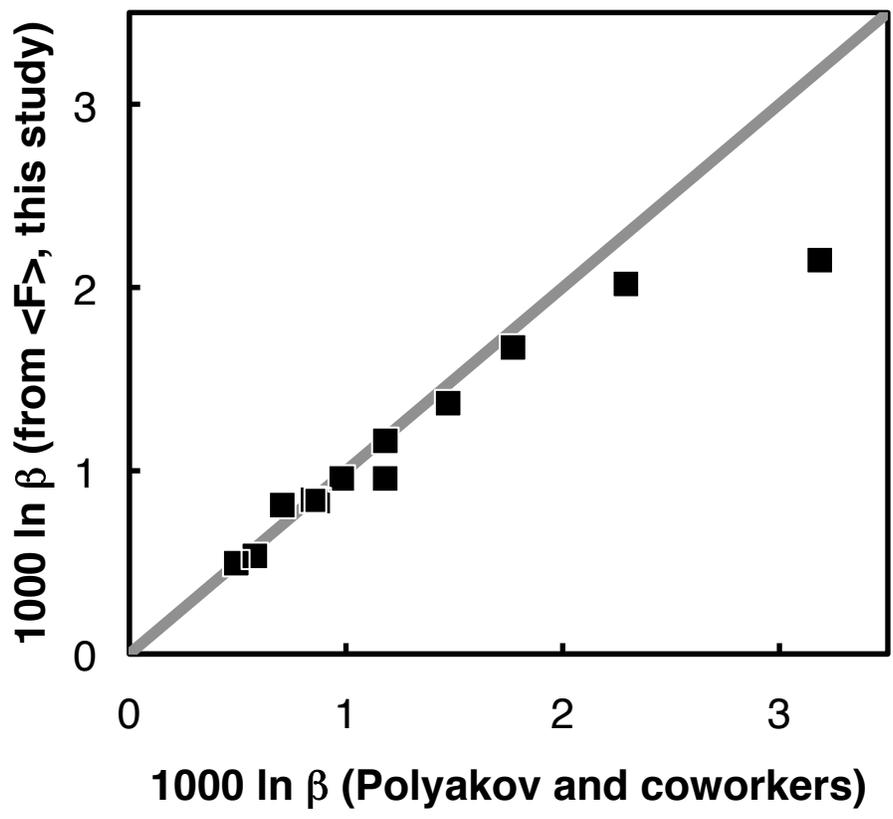

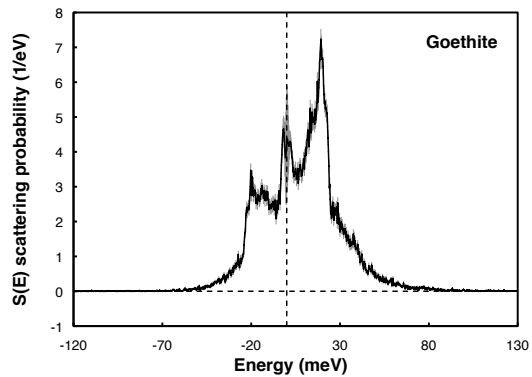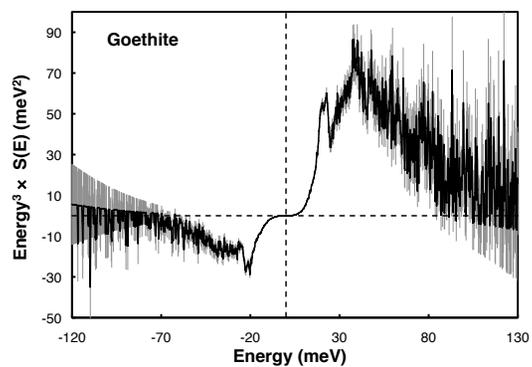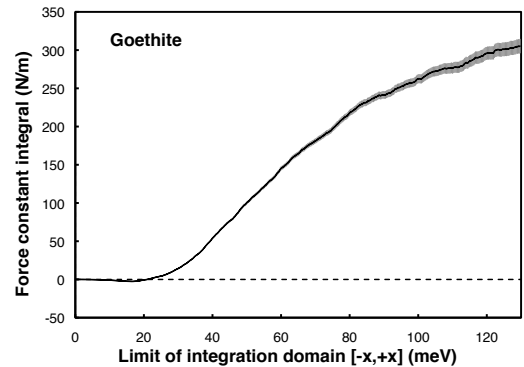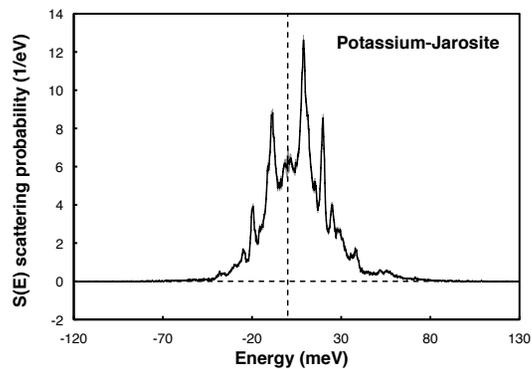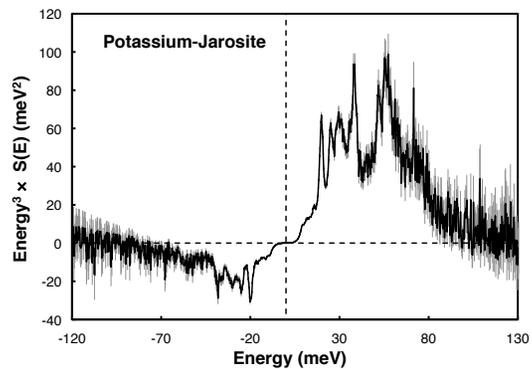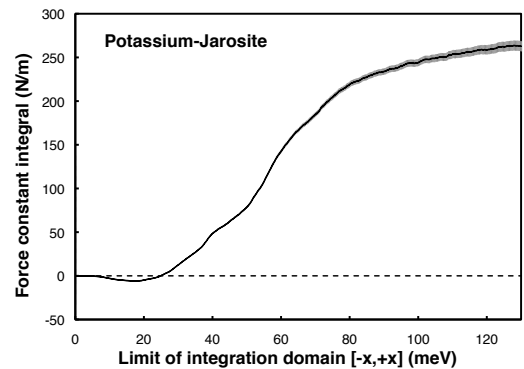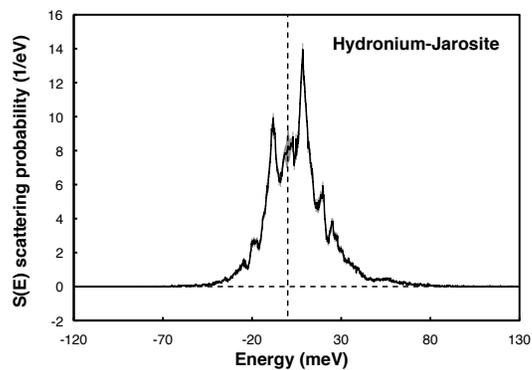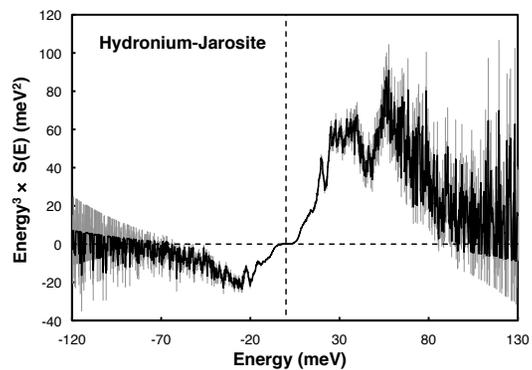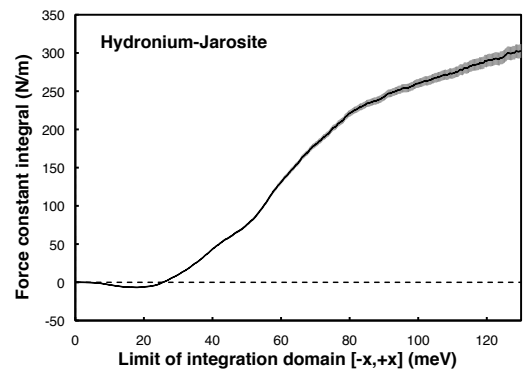

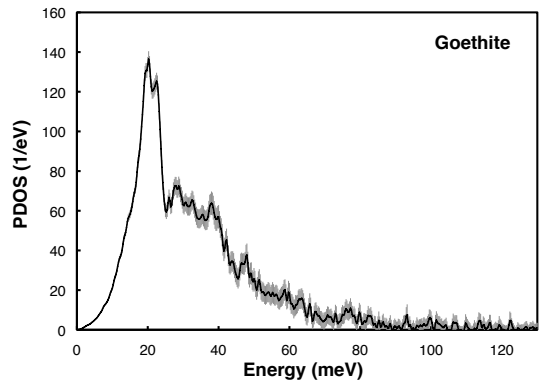 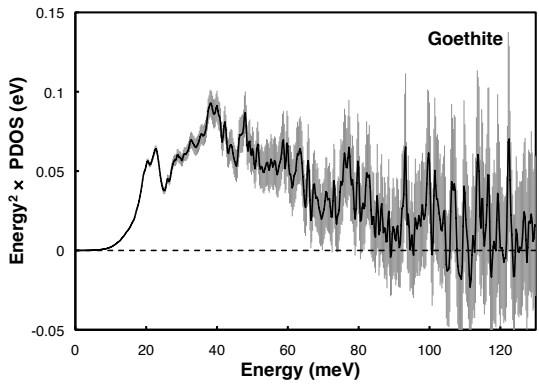 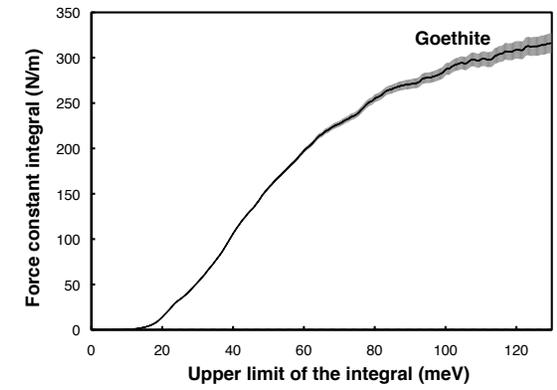
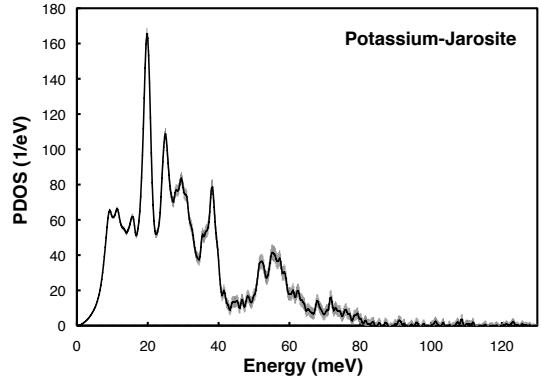 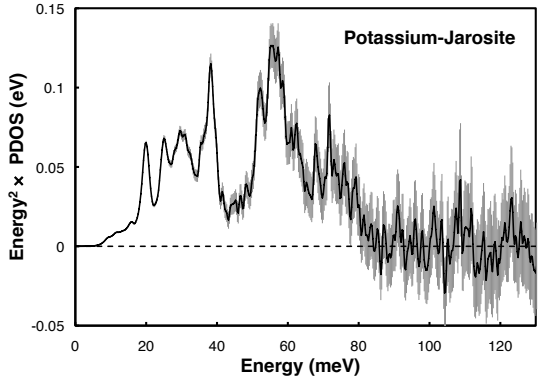 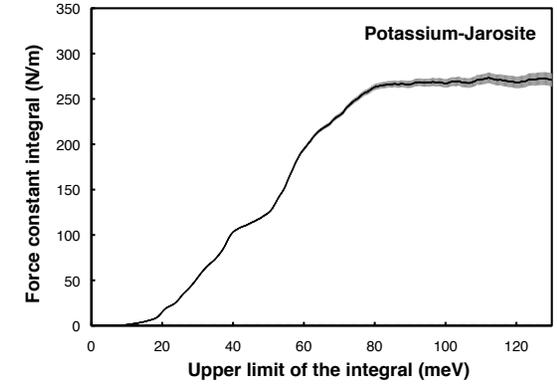
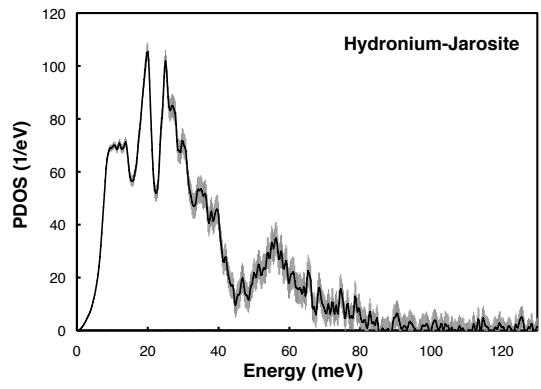 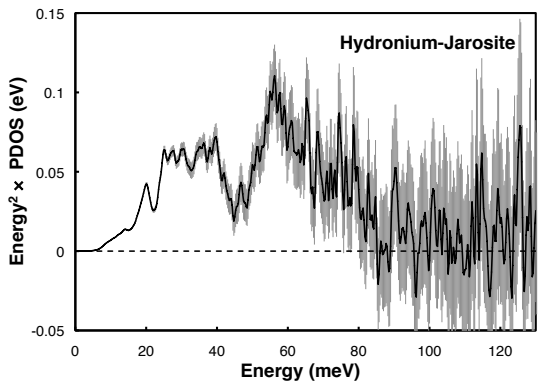 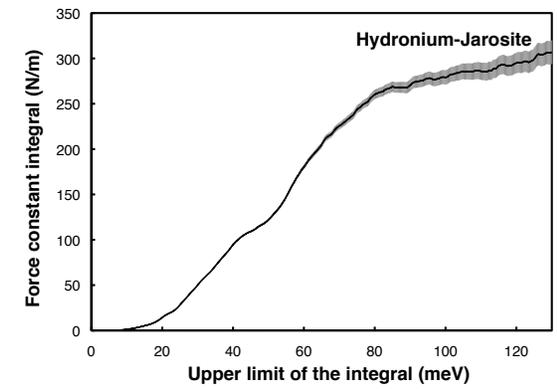

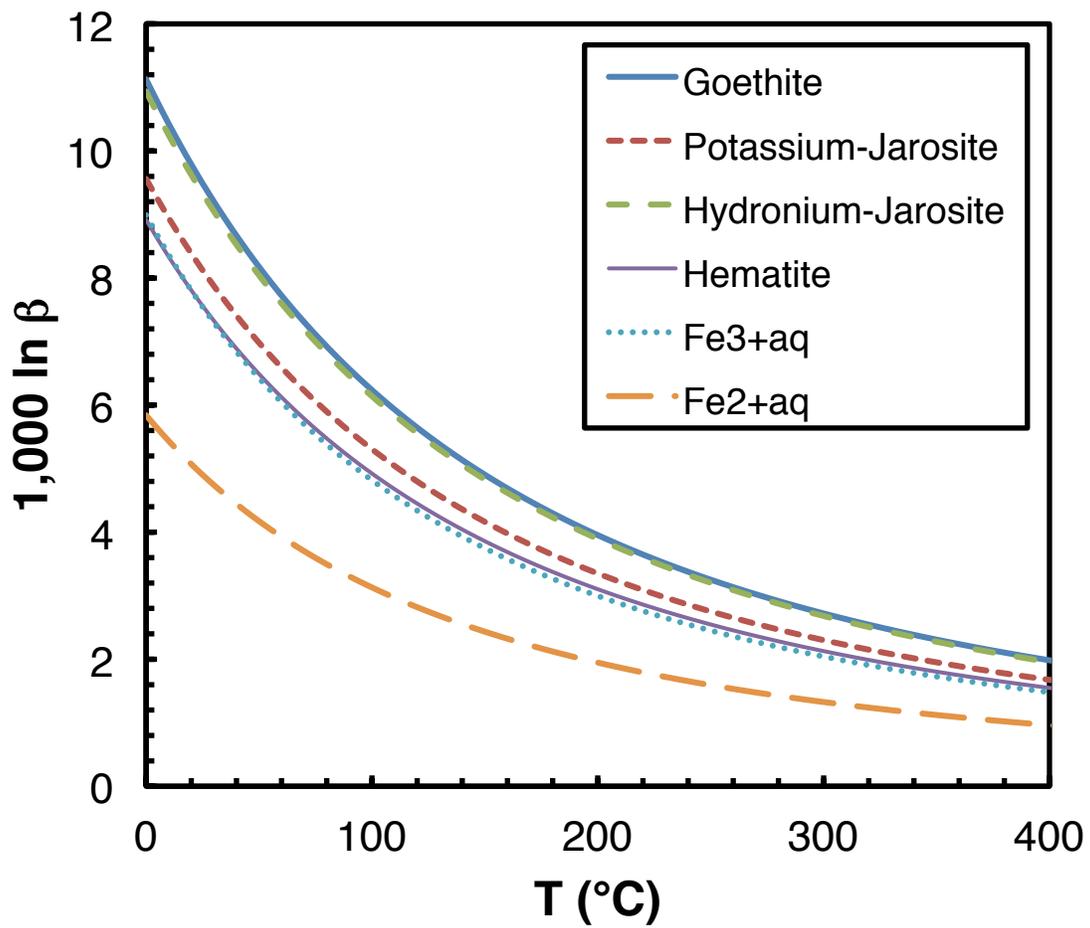

```mathematica
(*calculation of equilibrium isotopic fractionation from NRIXS data; Dauphas 2011*)
Av = 6.02214179 × 10^23; (*Avogadro number*)
m4 = 53.9396105 × 10^-3 / Av; m6 = 55.9349375 × 10^-3 / Av;
m7 = 56.9353940 × 10^-3 / Av; (*mass of a Fe atoms in kg*)
k = 1.3806503 × 10^-23 ; (*Boltzmann constant m2kgs-2*)
h = 6.626068 × 10^-34 ; (*Planck constant*)
hb = h / (2 * Pi);  (*hbar*)
Er = 1.956 * 0.001 * 1.60217646 × 10^-19;  (*57Fe recoil energy J*)

datafpath =
  "/Users/dauphasu/Desktop/Jarosite_Goethite_GCA2011/sourcedata/data_Feb11copy/";
datafname = "Goethite";

Te = 300.8; (*temperature of the experiments in K for calculation of flm from g(E)*)

datafdos = FileNameJoin[{datafpath, StringJoin[datafname, ".dos"]}];
(*density of states*)
datafpsn = FileNameJoin[{datafpath, StringJoin[datafname, ".psn"]}];
(*excitation function S*)
Print[datafdos, "  FOUND:   ", FileExistsQ[datafdos]]
Print[datafpsn, "  FOUND:   ", FileExistsQ[datafpsn]]

mydatag = Import[datafdos, "Table"];
mydataS = Import[datafpsn, "Table", HeaderLines → 1];

(*plotting of S(E) and g(E)*)
ListPlot[Transpose[{mydataS[[All, 1]], mydataS[[All, 2]]}],
 AxesLabel → {"E (meV)", "S(E) (1/eV)"}, Joined → True]
ListPlot[Transpose[{mydatag[[All, 1]], mydatag[[All, 2]]}],
 AxesLabel → {"E (meV)", "g(E) (1/eV)"}, Joined → True]

(*Excitation function S(E)*)
(*the first column in E in meV, the second is S(E) in 1/eV,
and the third in the error of S(E) in 1/eV*)
EnS = mydataS[[All, 1]] * 0.001 * 1.60217646 × 10^-19; (*mEV to J*)
SE = mydataS[[All, 2]] / (1.60217646 × 10^-19); (*1/eV to 1/J*)
eS = mydataS[[All, 3]] / (1.60217646 × 10^-19); (*1/eV to 1/J*)
(*DOS g(E)*)
(*the first column in E in meV, the second is g(E) in 1/eV,
and the third in the error of g(E) in 1/eV*)
Eng = mydatag[[All, 1]] * 0.001 * 1.60217646 × 10^-19; (*mEV to J*)
gE = mydatag[[All, 2]] / (3 * 1.60217646 × 10^-19); (*1/eV to 1/J*)
eg = mydatag[[All, 3]] / (3 * 1.60217646 × 10^-19); (*1/eV to 1/J*)

g = Interpolation[Transpose[{Eng, gE}], InterpolationOrder → 1];
(*interpolated function*)
S = Interpolation[Transpose[{EnS, SE}], InterpolationOrder → 1];
(*interpolated function*)

flm = Exp[-Er * NIntegrate[(g[x] / x) * Coth[x / (2 * k * Te)],
    {x, Min[Eng] + 0.5 * 0.001 * 1.60217646 × 10^-19, Max[Eng]}]]

(*calculation of the nth order moment of the excitation function*)
mS[n_] := NIntegrate[S[x] * (x - Er)^n, {x, Min[EnS], Max[EnS]}];
(*numerical integration done by Mathematica on interpolated function*)
(*numerical integration using the Trapezoidal rule*)
mSt[n_] := 0.5 * Sum[(EnS[[i + 1]] - EnS[[i]]) *
    (SE[[i + 1]] * (EnS[[i + 1]] - Er)^n + SE[[i]] * (EnS[[i]] - Er)^n), {i, 1, Length[EnS] - 1}];
```

```
(*error propagation in the integration using the Rectangule rule*)
emSr[n_] :=
  √Sum[((EnS[[i+1]] - EnS[[i]]) * ((eS[[i]]) * (EnS[[i]] - Er)^n))^2, {i, 1, Length[EnS] - 1}];
(*error propagation in the integration using the Trapezoidal rule*)
emSt[n_] := 0.5 * (((EnS[[2]] - EnS[[1]]) * eS[[1]] * (EnS[[1]] - Er)^n)^2 +
      Sum[((EnS[[i+1]] - EnS[[i-1]]) * eS[[i]] * (EnS[[i]] - Er)^n)^2,
        {i, 2, Length[EnS] - 1}] + ((EnS[[Length[EnS]]] - EnS[[Length[EnS] - 1]]) *
          eS[[Length[EnS]]] * (EnS[[Length[EnS]]] - Er)^n)^2)^0.5;
(*moments of S*)
S2 = mS[2]; S2t = mSt[2]; eS2r = emSr[2]; eS2t = emSt[2];
S3 = mS[3]; S3t = mSt[3]; eS3r = emSr[3]; eS3t = emSt[3];
S4 = mS[4]; S4t = mSt[4]; eS4r = emSr[4]; eS4t = emSt[4];
S5 = mS[5]; S5t = mSt[5]; eS5r = emSr[5]; eS5t = emSt[5];
S6 = mS[6]; S6t = mSt[6]; eS6r = emSr[6]; eS6t = emSt[6];
S7 = mS[7]; S7t = mSt[7]; eS7r = emSr[7]; eS7t = emSt[7];

Gs2 = S3 / Er;
eGs2 = eS3t / Er;
Gs4 = (S5 - 10 * S2 * S3) / Er;
eGs4 = (eS5t^2 + 10^2 * S2^2 * eS3t^2 + 10^2 * S3^2 * eS2t^2)^0.5 / Er;
Gs6 = (S7 + 210 * S2^2 * S3 - 35 * S3 * S4 - 21 * S2 * S5) / Er;
eGs6 =
  (eS7t^2 + 210^2 * S2^4 * eS3t^2 + 210^2 * S3^2 * 4 * S2^2 * eS2t^2 + 35^2 * S3^2 * eS4t^2 +
     35^2 * S4^2 * eS3t^2 + 21^2 * S2^2 * eS5t^2 + 21^2 * S5^2 * eS2t^2)^0.5 / Er;

(*force constant in N/m from S(E)*)
Fs = m7/(Er * hb^2) * S3;
eFs = m7/(Er * hb^2) * eS3t;

(*calculation of the nth order moments of the PDOS*)
mG[n_] := NIntegrate[g[x] × x^n, {x, Min[Eng], Max[Eng]}]
(*integration using the interpolated function*)
mGt[n_] :=
 0.5 * Sum[(Eng[[i+1]] - Eng[[i]]) * (gE[[i+1]] * (Eng[[i+1]])^n + gE[[i]] * (Eng[[i]])^n),
    {i, 1, Length[Eng] - 1}] (*numerical integration using the Trapezoidal method*)
emGr[n_] := √Sum[((Eng[[i+1]] - Eng[[i]]) * ((eg[[i]]) * (Eng[[i]])^n))^2,
    {i, 1, Length[Eng] - 1}]
    (*error propagation in the integration using the Rectangular method*)
emGt[n_] := 0.5 * (((Eng[[2]] - Eng[[1]]) * eg[[1]] * (Eng[[1]])^n)^2 +
      Sum[((Eng[[i+1]] - Eng[[i-1]]) * eg[[i]] * (Eng[[i]])^n)^2, {i, 2, Length[Eng] - 1}] +
      ((Eng[[Length[Eng]]] - Eng[[Length[Eng] - 1]]) *
          eg[[Length[Eng]]] * (Eng[[Length[Eng]]])^n)^2)^0.5
(*error propagation in the integration using the Trapezoidal method*)

(*2nd, 4th, and 6th orders*)
G2 = mG[2]; G2t = mGt[2]; eG2r = emGr[2]; eG2t = emGt[2];
G4 = mG[4]; G4t = mGt[4]; eG4r = emGr[4]; eG4t = emGt[4];
G6 = mG[6]; G6t = mGt[6]; eG6r = emGr[6]; eG6t = emGt[6];

(*calculation of the force constant and the coefficients of the polynomial*)
Fg = m7/hb^2 * G2; (*force constant from PDOS in N/m*)
eFg = m7/hb^2 * eG2t; (*uncertainty on force constant*)
```

```mathematica
Print["Force constant from PDOS= ", Fg,
 "±", eFg, "; from S(E)= ", Fs, "±", eFs, "  (N/m)"]
(*the following print out compares the 2n moment of g with that calculated from S*)
Print["G2/Gs2s: ", G2 / Gs2, "; G4/Gs4: ", G4 / Gs4, "; G6/Gs6: ", G6 / Gs6]
D1g = 1000 × (m7 (1 / m4 - 1 / m6)) / (8 × k^2) × G2; eD1g = 1000 × (m7 (1 / m4 - 1 / m6)) / (8 × k^2) × eG2t;
D2g = -1000 × (m7 (1 / m4 - 1 / m6)) / (480 × k^4) × G4; eD2g = 1000 × (m7 (1 / m4 - 1 / m6)) / (480 × k^4) × eG4t;
D3g = 1000 × (m7 (1 / m4 - 1 / m6)) / (20 160 × k^6) × G6; eD3g = 1000 × (m7 (1 / m4 - 1 / m6)) / (20 160 × k^6) × eG6t;
D1S = 1000 × (m7 (1 / m4 - 1 / m6)) / (8 × k^2) × Gs2; eD1S = 1000 × (m7 (1 / m4 - 1 / m6)) / (8 × k^2) × eGs2;
D2S = -1000 × (m7 (1 / m4 - 1 / m6)) / (480 × k^4) × Gs4; eD2S = 1000 × (m7 (1 / m4 - 1 / m6)) / (480 × k^4) × eGs4;
D3S = 1000 × (m7 (1 / m4 - 1 / m6)) / (20 160 × k^6) × Gs6; eD3S = 1000 × (m7 (1 / m4 - 1 / m6)) / (20 160 × k^6) × eGs6;

Print["from g(E): D₁=", D1g ± eD1g, "; D₂=", D2g ± eD2g, "; D₃=", D3g ± eD3g]
Print["from S(E): D₁=", D1S ± eD1S, "; D₂=", D2S ± eD2S, "; D₃=", D3S ± eD3S]

tmin = 0; tmax = 500; tstep = 10; (*temperature in celsius*)
(*calculates the beta values using the full
 integral for temperatures tmin to tmax in increments tstep*)
betag = Table[{i, 1000 × m7 (1 / m4 - 1 / m6) * 1.5 *
    (NIntegrate[g[x] × ((x / (k * (i + 273.15))) / (e^(x/(k*(i+273.15))) - 1) + 0.5 * x / (k * (i + 273.15)) - 1),
     {x, Min[Eng], Max[Eng]}])}, {i, tmin, tmax, tstep}];
betaS = Table[{i, D1S / (i + 273.15) ^ 2 + D2S / (i + 273.15) ^ 4 + D3S / (i + 273.15) ^ 6},
   {i, tmin, tmax, tstep}];

(*calculation of the coefficients in the polynomial 1000 ln beta=
 B1<F>/T^2+B2<F>^2/T^4 with <F> in N/m and T  in K*)
B1 = hb^2 * 1000 * (1 / m4 - 1 / m6) / (8 * k^2);
betares = betaS;
betares[[All, 2]] = betares[[All, 2]] - D1S / (betares[[All, 1]] + 273.15) ^ 2;
B2 = -Numerator[Fit[betares, {1 / (x + 273.15) ^ 4}, x]] / Fs^2;
Print["B₁=", B1, "; B₂=", B2]
Show[Plot[{B1 * Fs / (x + 273.15)^2 - B2 * Fs^2 / (x + 273.15)^4},
  {x, 0, 100}, AxesLabel → {"T (K)", "1000 ln β"}], ListPlot[betaS]]
Show[Plot[{{D1g / (x + 273.15)^2 + D2g / (x + 273.15)^4 + D3g / (x + 273.15)^6},
   {D1S / (x + 273.15)^2 + D2S / (x + 273.15)^4 + D3S / (x + 273.15)^6}},
  {x, 0, 100}, AxesLabel → {"T (K)", "1000 ln β"}], ListPlot[betaS]]
```

/Users/dauphasu/Desktop/Jarosite_Goethite_GCA2011/sourcedata/data_Feb11copy/Goethite.dos
    FOUND:    True

/Users/dauphasu/Desktop/Jarosite_Goethite_GCA2011/sourcedata/data_Feb11copy/Goethite.psn
    FOUND:    True

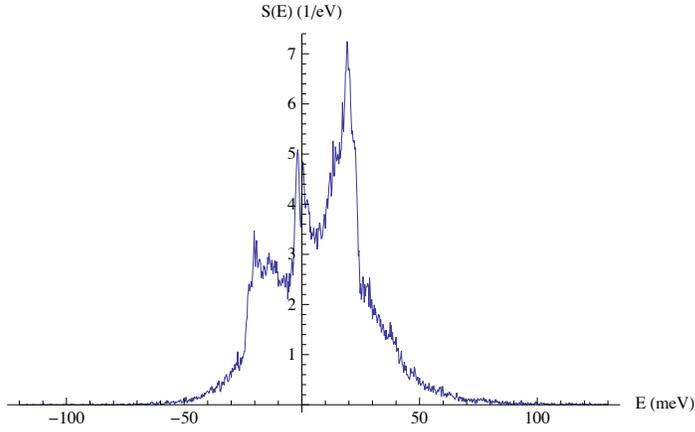

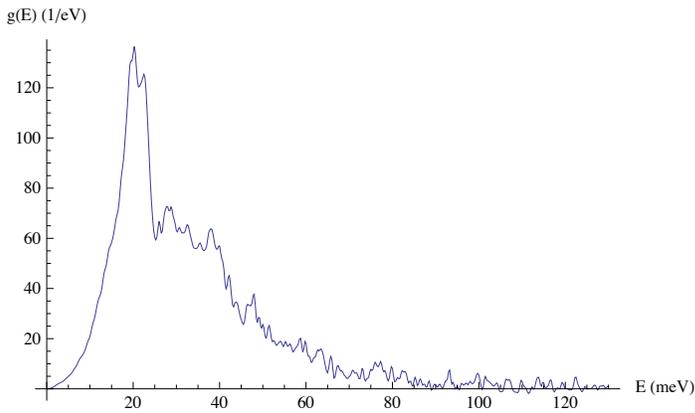

0.773692

Force constant from PDOS= 316.334±10.8674; from S(E)= 307.415±9.40995 (N/m)

G2/Gs2s: 1.02901; G4/Gs4: 1.10801; G6/Gs6: 1.41387

from g(E): $D_1$=918787. ± 31564.1; $D_2$= $-8.1491 \times 10^9 \pm 8.29987 \times 10^8$; $D_3$=$1.9636 \times 10^{14} \pm 3.59325 \times 10^{13}$

from S(E): $D_1$=892881. ± 27331.; $D_2$= $-7.35471 \times 10^9 \pm 6.92952 \times 10^8$; $D_3$=$1.38881 \times 10^{14} \pm 2.97415 \times 10^{13}$

$B_1$=2904.48; $B_2$=61951.3

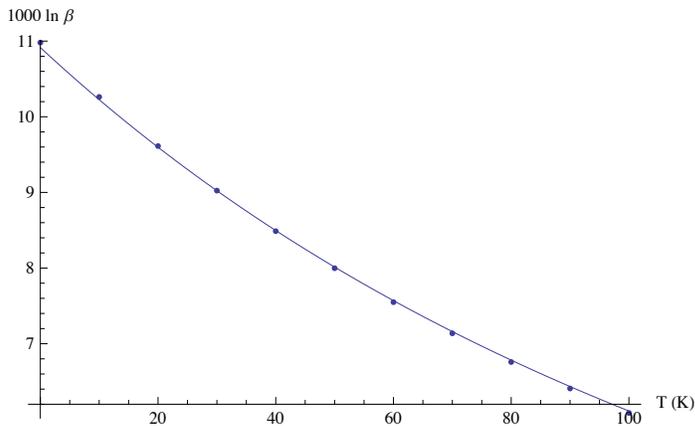

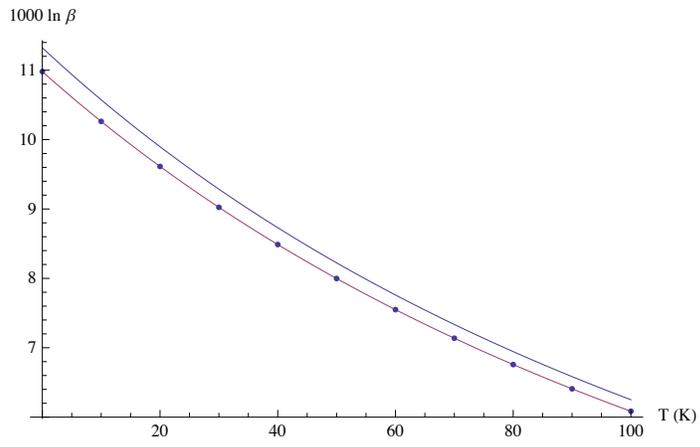